\documentclass[11pt,groupedaddress,showkeys]{revtex4-2}
\usepackage{float}
\usepackage{multirow}
\usepackage{amsmath}
\usepackage{amssymb}
\usepackage{graphicx}
\usepackage{stackrel}
\usepackage{natbib}
\usepackage[unicode=true]{hyperref}

\begin{document}
	\title{Probing the effects of primordial black holes on 21-cm EDGES signal along with interacting dark energy and dark matter - baryon scattering}
	
	\author{Ashadul Halder}
	\email{ashadul.halder@gmail.com}
	\affiliation{Department of Physics, St. Xavier's College, 30, Mother Teresa Sarani, Kolkata-700016, India.}
	
	\author{Madhurima Pandey}
	\email{madhurima0810@gmail.com}
	\affiliation{Theory Division, Saha Institute of Nuclear Physics, HBNI, 1/AF Bidhannagar, Kolkata-700064, India.\\
		Department of Physics, School of Applied Sciences and Humanities,  Haldia Institute of Technology, Haldia, West Bengal, 721657, India.}
	
	\begin{abstract}
		\begin{center}
			\large{\bf Abstract}
		\end{center}
		
		21-cm radio signal has emerged as an important probe in investigating the dark age of the Universe (recombination to reionization). In the current analysis, we explore the combined effects of primordial black holes (PBH), cooling off of the baryonic matter due to dark matter (DM) - baryon collisions and interaction of dark matter - dark energy (DE) fluid on the 21-cm brightness temperature. The variation of brightness temperature shows remarkable dependence on DM mass ($m_{\chi}$) and the dark matter - baryon scattering cross-section ($\sigma_0$). Bounds in $m_{\chi}$ - $\sigma_0$ parameter space are obtained for different possible PBH masses and for different interacting dark energy (IDE) models. These bounds are estimated based on the observed excess ($-500^{+200}_{-500}$ mK) of 21-cm brightness temperature by EDGES experiment. Eventually, bounds on PBH mass is also obtained for different values of dark matter mass and for different IDE model coupling parameters. The compatibility of the constraints of the IDE models, in the estimated bounds are also addressed. 
		
	\end{abstract}
	\maketitle


\section{Introduction}
The 21-cm cosmology is turning out to be a promising tool in understanding 
the dynamics of the early Universe. The redshifted signature of the 21-cm 
neutral hydrogen spectrum opens up a new window to understand the process 
of reionization and the factors in the early Universe influencing the same. 
Thus, study of the 21-cm line in reionization era helps in understanding 
several cosmological and astrophysical processes that might have 
contributed to the physics of the early Universe.

The 21-cm ($\sim 1.42$ GHz) hyperfine spectrum is originated due to the 
transition between two spin states ($s=0$ and 1) of the neutral H atoms. 
The hydrogen occupies around $75\%$ of the entire baryonic mass of the Universe. 
The corresponding spin temperature indicates the population of hydrogen 
atoms with different energy states. 
The ``Experiment to Detect the Global Epoch of Reionization Signature'' 
(EDGES) \cite{edges} reported 21-cm absorption spectra at 
the cosmic dawn era ($14<z<20$) and predicted the 21-cm brightness 
temperature to be $-500^{+200}_{-500}$ mK with $99\%$ confidence level (C.L.). 
This measured brightness temperature is lower than the expected value.

The 21-cm brightness temperature $T_{21}$ is related to the 
temperature difference $T_s - T_\gamma$, where
$T_s$ is the spin temperature and the $T_\gamma$ is the background temperature
(CMB temperature). The observed additional cooling of $T_{21}$ by EDGES
experiment can be
realized by either enhancing the background temperature $T_\gamma$ or
by lowering the matter temperature which is equal to $T_s$ at that epoch.
Dark matter (DM) interactions such as the scattering of dark matter off baryons, dark matter annihilation or decay can inject energy into
the background resulting in the rise in background temperature.
There can be other processes such as possible dark matter - dark energy 
interaction, which can induce the larger than expected
difference of $T_s$ and $T_\gamma$. In this work, both these 
possibilities are explored. In addition, the possibility that the 
evaporation of Primordial
Black Holes (PBHs) injecting more energy into the system is also addressed in 
this work. 

PBHs \cite{khlopov_1,khlopov_2,khlopov_3,juan} are believed to be formed 
during the
radiation dominated era. PBHs forms due to the collapse of an overdensity
region characterized by the
size of the region which should be greater than the Jeans length $R_j$, 
where $R_j = \sqrt{\displaystyle\frac {1} {3G \rho}}$. Also the condition of the
PBH formation is $\delta_{\rm min} \leq \delta \leq \delta_{\rm max}$,
where $\delta$ is the density contrast. The maximum and minimum density
contrast $\delta_{\rm max}$ and $\delta_{\rm min}$ respectively are governed
by the value of $\delta \rho$, where
the density $\rho = \rho_c + \delta \rho$, $\rho_c$ being the
critical density for collapse and $\delta_{\rm min}$ is the threshold of PBH 
formation. Moreover, there are several mechanisms describing the formation 
of PBHs \cite{stpbh_1,stpbh_2,stpbh_3,stpbh_4,stpbh_5,stpbh_6,stpbh_7,fsc_1,fsc_2,fsc_3,ccs_1,ccs_2,ccs_3}. In the work of \cite{BH_21cm_1}, the masses of the PBHs 
$\geq 10^{15}$ g are adopted. But 
in the present work, we investigate the 21-cm signal with the 
PBH masses $10^{13} 
\leq \mathcal{M}_{\rm BH}\leq 10^{14}$ g. This range of PBH mass is also 
considered in \cite{BH_21cm_2}.

As mentioned earlier in this work, we address the possible influences of the 
factors namely 
DM - baryon scattering, possible DM - DE interaction \cite{idem0,idem1,idem2,idem3,Kumar_2017,Kumar_2019} and the evaporation of 
PBH \cite{BH_21cm_0,BH_21cm_1,BH_21cm_2,BH_21cm_4,BH_21cm_5,PhysRevD.98.023503,10.1093/mnras/stt1493} simultaneously 
on 21-cm EDGES signal in the epoch of ignition of first star. 
The heating effect by the PBHs are assumed to be contributed by 
Hawking Radiation from PBH only. For the 
DM - DE interaction three interacting dark energy (IDE) models 
given in \cite{Li} are chosen.	
The influence of DM - DE interaction on 21-cm signal has been discussed earlier 
in \cite{Li,upala}. Moreover, DM-DE interaction is discussed in several context such as in addressing the cosmological coincidence problem \cite{Wang_2016}, Hubble tension \cite{PhysRevD.96.043503}, Large Scale Structure formation \cite{Farrar_2004} etc.

The general form of the velocity dependent cross-section is given by $\bar{\sigma}=\sigma_0 (v/c)^n$, where the index $n$ depends on different physical dark matter processes and $c$ is the velocity of light in space (in natural unit $\bar{\sigma}=\sigma_0 v^n$). In the case of DM with magnetic and/or electric dipole moment $n=+2,-2$ are considered. $n=2,1,0,-1$ are applicable for scattering in presence of Yukawa potential \cite{yukawa}, $n=-4$ is attributed for millicharged DM \cite{mcharge1,mcharge2}. In Ref.~\cite{dvorkin2020cosmology} the nature of the DM-baryon cross-section is discussed for a wide mass range of dark matter. Similar investigations are also carried out in Ref.~\cite{Nadler_2019,Bhoonah_2018,Kovetz_2018,Mack_2007}. In the present work, the dark matter - baryon interaction cross-section ($\bar{\sigma}$) is parameterized as $\bar{\sigma}=\sigma_0 v^{-4}$ \cite{munoz,upala,rennan_3GeV}. 
The term $\sigma_0$ is the dark matter scalar scattering cross-section with baryons (of the type $\alpha_q \bar{\chi}\chi \bar{q}q$ for dark dark matter particle $\chi$ with coupling $\alpha_q$). It may be mentioned in some earlier works \cite{Bhoonah_2018,Kovetz_2018} millicharged dark matter is considered. But here, we assume a particle dark matter candidate and adopt value of $\sigma_0 \sim 10^{-41} \rm{cm^2}$ consistent with the scalar cross-section bound obtained from ongoing direct dark matter search experiments (extrapolating the allowed region for $0.1$ GeV$\leq m_{\chi} \leq 3$ GeV from recent experiments \cite{xenon1t,lux,pandax2}) in the mass range discussed in this work.
Several recent investigations on EDGES 21-cm signal also suggest the similar velocity dependence ($n=-4$) of the cross-section \cite{munoz,Mahdawi_2018,rennan_3GeV}. Moreover, $n=-4$ is chosen in many dark matter related cases namely hadronically interacting DM, millicharge DM, the Baryon Acoustic Oscillations (BAO) signal etc.


The paper is organized as follows. In Section~\ref{sec:DMDE}, we 
address the interaction between dark matter and dark energy and its 
effect in cosmic evolution. 
Section~\ref{sec:PBH} deals with the injection by the PBHs 
in the form of Hawking 
radiation. In Section~\ref{sec:T_evol}, the formalism of evolutions of various 
temperatures such as $T_\chi, T_b$ (DM temperature, baryon temperature) along 
with the effect of PBH evaporation are described. Section~\ref{sec:21cm} 
describes the formalism for 21-cm absorption line. Calculations and results 
are shown in Section~\ref{sec:result}. Finally in 
Section~\ref{sec:conc}, some concluding remarks are given.

\section{Dark Matter - Dark Energy Interaction} \label{sec:DMDE} 
The DM - DE interaction may have a profound effect in the universal dynamics 
and hence on the optical depth and spin temperature of the 21-cm transition.
In standard cosmological model, the density parameters of dark matter 
($\Omega_{\chi}$) and dark energy ($\Omega_{\rm de}$) are assumed to be evolved 
as $\Omega_{\chi,0}(1+z)^3$ and $\Omega_{\rm de,0}(1+z)^{3(1+\omega)}$ where,
$\Omega_{\chi,0}$ and $\Omega_{\rm de,0}$ are the respective density parameters at $z=0$ 
and $\omega$ is the equation of state (EOS) parameter of dark energy. However, 
if the interaction between dark matter and dark energy is taken into account, 
the evolution of dark matter and dark energy densities 
take the forms \cite{Li},
\begin{equation}
	(1+z) H(z) \dfrac{{\rm d} \rho_{\chi}}{{\rm d} z}-3H(z)\rho_{\chi}  = -\mathcal{Q} 
	\label{eq:rho_chi}
\end{equation}
\begin{equation}
	(1+z) H(z) \dfrac{{\rm d} \rho_{\rm de}}{{\rm d} z} - 3 H(z) (1+ \omega) \rho_{de} = \mathcal{Q}
	\label{eq:rho_de}
\end{equation}
where $\mathcal{Q}$ denotes the energy transfer between dark matter and dark 
energy due to DM-DE interaction. In the present work, we consider three benchmark 
models in order to investigate the effect of DM - DE interaction in the brightness temperature. The energy transfer expressions of those benchmark models 
are described below \cite{model1,model2,model3,model4}.
\begin{center}
	\begin{tabular}{ll}
		Model-I \hspace{5mm} & $\mathcal{Q}=3 \lambda H(z) \rho_{\rm de}$\\
		Model-II & $\mathcal{Q}=3 \lambda H(z) \rho_{\chi}$\\
		Model-III &$\mathcal{Q}=3 \lambda H(z) (\rho_{\rm de} +\rho_{\chi})$
	\end{tabular}
\end{center}
Here, $\lambda$ is the coupling parameter, which determine the strength of the dark matter - dark energy interaction. The stability conditions for each of the models are described in Table~\ref{tab:stability}.
Several phenomenological studies have been carried out with observational 
data of PLANCK, Supernova Ia (SNIa) Baryon Acoustic Oscillation (BAO) 
\cite{model3, model4, model_benchmark1, model_benchmark2, model_benchmark3,
	model_benchmark4, model_benchmark5, model_benchmark6} yielding the 
constraints for different models (in Table~\ref{tab:constraints}). 
It is to be mentioned that, all the IDE models discussed in this section are independent of the dark matter - baryon interaction.
\begin{table}
	\centering
	\caption{\label{tab:stability} Stability conditions of the model parameters for different IDE models}
	\begin{tabular}{lccr}
		\hline
		Model & $\mathcal{Q}$ & EOS of dark energy & Constraints\\
		\hline
		I & 3 $\lambda H(z) \rho_{\rm de} $ & $\omega<-1$ & $\lambda<- 2 \omega \Omega_{\chi}$\\
		II & 3 $\lambda H(z) \rho_{\chi} $ & $\omega<-1$ & $0<\lambda<-\omega/4$\\ 
		III & 3 $\lambda H(z) (\rho_{\rm de} + \rho_{\chi}) $ & $\omega<-1$ & $0<\lambda<-\omega/4$\\
		\hline
	\end{tabular}
	
\end{table}

\begin{table}
	\centering
	\caption{\label{tab:constraints} Constraints of the different IDE models}
	\begin{tabular}{lccc}
		\hline
		Model & $\omega$ & $\lambda$ & $H_0$\\
		\hline
		
		$3 \lambda H \rho_{\rm de}$ & $-1.088^{+0.0651}_{-0.0448}$ & $0.05219^{+0.0349}_{-0.0355}$ & $68.35^{+1.47}_{-1.46}$\\
		
		$3 \lambda H \rho_{\chi}$ & $-1.1041^{+0.0467}_{-0.0292}$ & $0.0007127^{+0.000256}_{-0.000633}$ & $68.91^{+0.875}_{-0.997}$\\
		
		$3 \lambda H (\rho_{\rm de}+\rho_{\chi})$ & $-1.105^{+0.0468}_{-0.0288}$ & $0.000735^{+0.000254}_{-0.000679}$ & $68.88^{+0.854}_{-0.97}$\\
		\hline
	\end{tabular}
\end{table}

\section{Effect of Primordial Black Hole} \label{sec:PBH} 
The energy injection of PBHs in the form of Hawking radiation \cite{BH_F} can be 
a possible source for heating up of the medium before 
the reionization. It has been shown by \cite{BH_21cm_1} that, in 21-cm scenario, 
the Hawking radiation is equally significant as that of the DM decay.

The mass evaporation rate due to Hawking radiation can be expressed as
\begin{equation}
	\dfrac{{\rm d}M_{\rm{BH}}}{{\rm d}t} \approx -5.34\times10^{25} \left(\sum_{i} \mathcal{F}_i\right) \left(\dfrac{M_{\rm{BH}}}{\rm g}\right)^{-2} \,\,\rm{g/sec}
	\label{eq:PBH}
\end{equation}
where, $M_{\rm{BH}}$ is the mass of black hole and $\sum_{i} \mathcal{F}_i$ is the sum over all fraction of evaporation, defined as \cite{BH_F},
\begin{eqnarray}
	\sum_{i} \mathcal{F}_i&=&1.569+0.569 \exp 
	\left(-\frac{0.0234}{T_{\rm{BH}}} \right)+
	3.414\exp \left(-\frac{0.066}{T_{\rm{BH}}} \right)\nonumber\\
	&&+ 1.707\exp \left(-\frac{0.11}{T_{\rm{BH}}} \right)+ 
	0.569\exp \left(-\frac{0.394}{T_{\rm{BH}}} \right)\nonumber\\
	&&+1.707\exp \left(-\frac{0.413}{T_{\rm{BH}}} \right)+ 
	1.707\exp \left(-\frac{1.17}{T_{\rm{BH}}} \right)\nonumber\\
	&&+1.707\exp \left(-\frac{22}{T_{\rm{BH}}} \right)+
	0.963\exp \left(-\frac{0.1}{T_{\rm{BH}}}\right)
\end{eqnarray}
In the above expression, $T_{\rm BH}$ represents the temperature of the black hole 
given by, $T_{\rm{BH}}=1.05753 \times \left(M_{\rm{BH}}/10^{13} 
{\rm g}\right)^{-1}$. 
In the case of massive black holes, only the contributions of photon and electron channels are significant. However, in the present work, the mass range of the PBHs are considered to be $\sim 10^{14}$--$10^{15}$ g. The temperature of such PBHs are substantially high to radiate in the form of pions, muons, quarks and gluons \cite{PhysRevD.41.3052,PhysRevD.94.044029,BH_21cm_2}. As a consequence, besides the $\gamma$ and electron channels, other channel also contribute remarkably to the IGM heating by producing photons, electrons and positrons via subsequent cascade decay \cite{PhysRevD.41.3052,chen,PhysRevD.81.104019}. The energy injection rate per unit volume due to PBHs is given by \cite{BH_21cm_2},
\begin{equation}
	\left.\dfrac{{\rm d} E}{{\rm d}V {\rm d}t}\right|_{\rm{BH}}=-\dfrac{{\rm d} M_{\rm{BH}}}{{\rm d} t} n_{\rm BH}(z)
\end{equation}
where, $n_{\rm{BH}}(z)$ is the number density of black hole at redshift $z$. which can be expressed as a function of 
cosmological redshift ($z$) and initial mass fraction of primordial black holes 
($\beta_{\rm BH}$), as,
\begin{eqnarray}
	n_{\rm{BH}}(z)&=&\beta_{\rm BH}\left(\dfrac{1+z}{1+z_{\rm eq}}\right)^3 \dfrac{\rho_{\rm c,eq}}{\mathcal{M}_{\rm BH}} \left(\dfrac{\mathcal{M}_{\rm H,eq}}{\mathcal{M}_{\rm H}}\right)^{1/2} \left(\dfrac{g^i_{\star}}{g^{\rm eq}_{\star}}\right)^{1/12}\nonumber\\
	&\approx&1.46 \times 10^{-4}\beta_{\rm BH} \left(1+z\right)^3 \left(\dfrac{\mathcal{M}_{\rm BH}}{\rm g}\right)^{-3/2} {\rm cm^{-3}}.
\end{eqnarray}

\section{Temperature Evolution} \label{sec:T_evol} 
In this Section, the formalism of evolution of 
baryon temperature ($T_b$) and the dark matter temperature ($T_{\chi}$) 
with redshift $z$ is discussed. As mentioned earlier, we have considered three effects namely DM - baryon scattering, DM - DE interaction and evaporation of PBH in temperature evolution of baryon and dark matter and finally compute the 21-cm brightness temperature ($T_{21}$). 
The effect of dark matter - baryon scattering has earlier been addressed 
in the context of 21-cm signal by \cite{munoz}. 
More recently the effect of DM - DE interaction is also included along 
with DM - baryon scattering by \cite{upala}. 
In the present work, in addition, the effects of PBH evaporation are also included along with DM - baryon  
scattering and DM - DE interaction in the evolution equations of $T_{\chi}$ and 
$T_{b}$. With all these, the temperature evolution of $\chi$ and baryon $b$ can be written as
\begin{equation}
	(1+z)\frac{{\rm d} T_\chi}{{\rm d} z} = 2 T_\chi - \frac{2 \dot{Q}_\chi}{3 H(z)}-\frac{1}{n_\chi}\frac{2 \mathcal{Q}}{3 H(z)}, 
	\label{eq:T_chi}
\end{equation}
\begin{equation}
	(1+z)\frac{{\rm d} T_b}{{\rm d} z} = 2 T_b + \frac{\Gamma_c}{H(z)}
	(T_b - T_\gamma)-\frac{2 \dot{Q}_b}{ 3 H(z)}-\mathcal{J}_{\rm BH}. 
	\label{eq:T_b}
\end{equation}
where, the last term of Eq.~\ref{eq:T_chi} indicates the effects of dark 
matter - dark energy interaction (see Section~\ref{sec:DMDE}) and the last term 
of Eq.~\ref{eq:T_b} 
represents the contribution of PBHs in the form of Hawking radiation 
\cite{BH_21cm_2} given by,
\begin{equation}
	\mathcal{J}_{\rm BH}=\frac{2}{3 k_B H(z)}
	\frac{K_{\rm BH}}{1+f_{\rm He}+x_e}.
	\label{JBH}
\end{equation}
In Eq.~\ref{eq:T_b}, $T_{\gamma}$ 
($T_{\gamma}=2.725 (1+z)$ K) is the CMB temperature and $\Gamma_c$ 
($\Gamma_c=\frac{8\sigma_T a_r T^4_{\gamma}x_e}{3(1+f_{\rm He}+x_e)m_e c}$) 
describes the Compton interaction rate ($\sigma_T$ and $a_r$ are the 
Thomson scattering cross-section and radiation constant respectively). The 
quantities $f_{\rm He}$ and $x_e$ are the fractional abundance of He and ionization fraction 
respectively. The heating rates $\dot{Q}_b$ and $\dot{Q}_{\chi}$ are 
estimated as described in \cite{munoz} 
($b$ and $\chi$ represent baryon and DM respectively) which depends on the drag 
term $V_{\chi b}$.

The ionization fraction $x_e$ ($=n_e/n_H$, where $n_e$ and $n_H$ are the number density of free electron and hydrogen respectively) is an important quantity in estimating 
thermal evolution. It also influences $T_b$ and $T_{\gamma}$ simultaneously. This evolution is given by,	
\begin{equation}
	\frac{{\rm d} x_e}{{\rm d} z} = \frac{1}{(1+z)\,H(z)}\left[I_{\rm Re}(z)-
	I_{\rm Ion}(z)-I_{\rm BH}(z)\right], 
	\label{eq:xe}
\end{equation}
where $I_{\rm Re}(z)$ and $I_{\rm Ion}(z)$ are the standard recombination rate 
and standard ionization rate respectively. The combined effect of these two 
coefficients is described as,
\begin{equation}
	I_{\rm Re}(z)-I_{\rm Ion}(z) = C_P\left(n_H \alpha_B x_e^2-4(1-x_e)\beta_B 
	e^{-\frac{3 E_0}{4 k_B T_{\gamma}}}\right).
	\label{eq:xe_comp}
\end{equation}
In Eq.~\ref{eq:xe_comp}, $C_P$ is the Peebles C factor \cite{peeble,hyrec11}, $E_0=13.6$ eV, $\alpha_B$ and $\beta_B$ are the case B recombination and ionization coefficients respectively.

The expression for $\alpha_B$ (in $\rm{m^3}s^{-1}$) as a function of 
temperature, can be obtained by data fitting as obtained in the work 
of \cite{pequignot}. The fitted expression of $\alpha_B$ with parameters 
$a=4.309$, $b=-0.6166$, $c=0.6703$, $d=0.5300$, $F=1.14$ is given 
by \cite{pequignot,BH_21cm_5},
\begin{equation}
	\alpha_B=10^{-19} F\left(\frac{a t^b}{1+c t^d}\right), \label{alphaB}
\end{equation}
where, $t$ represents the temperature in $10^4$K 
\cite{hummer,pequignot,seager}. The expression for photoionization coefficient ($\beta_B$) 
(in term of $\alpha_B$) \cite{seager,BH_21cm_5} is,
\begin{equation}
	\beta_B=\alpha_B \left(\frac{2 \pi \mu_e k_B T_{\gamma}}{h^2}\right)^{3/2} 
	\exp\left(-\frac{h \nu_{2s}}{k_B T_{\gamma}}\right), \label{betaB}
\end{equation}
where, $\mu_e$ is the reduced mass of electron-proton system and $\nu_{2s}$ 
denotes the frequency for $2s\rightarrow 1s$ transition. The Peebles C factor reads as \cite{hyrec11},
\begin{equation}
	C_P=\dfrac{\frac{3}{4}R_{\rm Ly\alpha}+\frac{1}{4}\Lambda_{2s1s}}{\beta_B+
		\frac{3}{4}R_{\rm Ly\alpha}+\frac{1}{4}\Lambda_{2s,1s}}. \label{peeblec}
\end{equation}		
In the above, $R_{\rm Ly\alpha}$ represents the rate of escape of Lyman-$\alpha$ (Ly$\alpha$) photons 
$$
R_{\rm Ly\alpha}=8\pi H/\left(3 n_H (1-x_e)\lambda_{\rm Ly\alpha}^3\right)
$$ 
and $\Lambda_{2s,1s}\approx 8.22 \rm{s^{-1}}$ \cite{hyrec11}. 

In Eqs.~\ref{JBH} and \ref{eq:xe}, the parameters $K_{\rm BH}$ and $I_{\rm BH}$ are described as,
\begin{equation}
	K_{\rm BH}=\chi_{h} f(z) \frac{1}{n_b} \times \left.\dfrac{{\rm d} E}{{\rm d}V {\rm d}t}\right|_{\rm{BH}}. \label{KBH}
\end{equation}
\begin{equation}
	I_{\rm BH}=\chi_{i} f(z) \frac{1}{n_b} \frac{1}{E_0}\times \left.
	\dfrac{{\rm d} E}{{\rm d}V {\rm d}t}\right|_{\rm{BH}}, \label{IBH}
\end{equation}
where $\chi_{i}=(1-x_e)/3$ and $\chi_{h}=(1+2x_e)/3$ are the fraction of the energy deposited in the form of ionization and heating respectively \cite{BH_21cm_2,chen,BH_21cm_4,PhysRevD.76.061301,Furlanetto:2006wp}. The factor $f(z)$ is the total fraction of the injected energy deposited into the IGM at redshift $z$ \cite{corr_equs,fcz001,fcz002,fcz003,fcz004}.   

The evolution of relative velocity between the baryonic matter and dark matter 
$(V_{\chi b}\equiv V_{\chi}-V_{b})$ with 
redshift, as discussed in \cite{munoz} play an important role in this 
formalism. This is related to the drag term $D(V_{\chi b})$, between $\chi$ and $b$, as
\begin{equation}
	\frac{{\rm d} V_{\chi b}}{{\rm d} z} = \frac{V_{\chi b}}{1+z}+
	\frac{D(V_{\chi b})}{(1+z) H(z)}\,\,, \label{eq:V_chib}
\end{equation}
with initial condition $V_{\chi b} = 29$ km/s. In the above equation, 
the drag term $D(V_{\chi b})$ is defined as, 
\begin{equation}
	D(V_{\chi b})=\dfrac{{\rm d}(V_{\chi b})}{{\rm d}t}=\dfrac{\rho_m 
		\sigma_0}{m_b + m_{\chi}} \dfrac{1}{V^2_{\chi b}} F(r)
	\label{eq:dvchib}
\end{equation}
In Eq. \ref{eq:dvchib}, $F(r)={\rm erf}\left(r/\sqrt{2}\right)-\sqrt{2/\pi}r e^{-r^2/2}$, where $r$ is defined as $r=V_{\chi b}/u_{\rm th}$ and $u_{\rm th}=\sqrt{T_b/m_b+T_{\chi}/m_{\chi}}$ is the variance of the thermal relative motion of dark matter - baryon fluid and $\sigma_0$ is the dark matter - baryon scattering cross-section while $\sigma_{41}$ is the same in units of $10^{-41} {\rm cm^2}$.

\section{21-cm Cosmology} \label{sec:21cm}
As mentioned earlier, the 21-cm line is originated due to the transition of electrons between the 
triplet and singlet states of the hydrogen atom (spin 0 and spin 1). 
The intensity of the 21-cm line is represented by the brightness temperature 
($T_{21}$) which depends on optical depth ($\tau$) and hence the Hubble 
parameter ($H(z)$). The variation of brightness temperature of the 21-cm 
hydrogen spectrum with redshift ($z$) is given by,
\begin{equation}
	T_{21}=\dfrac{T_s-T_{\gamma}}{1+z}\left(1-e^{-\tau}\right)\approxeq
	\dfrac{T_s-T_{\gamma}}{1+z} \tau,
	\label{eq:t21}
\end{equation}
where, $T_s$ and $T_{\gamma}$ are spin temperature and CMB temperature 
respectively at redshift $z$. As the numerical values of $\tau$ for 
different $z$ are very small, we use the approximation in 
the above equation (Eq.~\ref{eq:t21}). The optical depth ($\tau$) is 
given by,
\begin{equation}
	\tau = \dfrac{3}{32 \pi}\dfrac{T_{\star}}{T_s}n_{\rm HI} \lambda_{21}^3\dfrac{A_{10}}{H(z)+(1+z)\delta_r v_r}
	\label{eq:tau}
\end{equation}
where, $\lambda_{21}$ ($\approx21$ cm) is the 21-cm wavelength, $T_{\star}$ ($=hc/k_B 
\lambda_{21}=0.068$ K) is the 21-cm temperature, $A_{10}$ ($=2.85\times 10^{-15}\,
{\rm s^{-1}}$) is the Einstein coefficient \cite{yacine} and $\delta_r v_r$ 
represents the  peculiar velocity gradient.

The spin temperature $T_s$ describes the ratio $n_1/n_0$, where $n_1$ 
and $n_0$ are excited state and ground state neutral hydrogen number 
densities respectively, given by 
$n_1/n_0=3 \exp{-T_{\star}/T_s}$. 
In equilibrium $T_s$ is given by
\begin{equation}
	T_s = \dfrac{T_{\gamma}+y_c T_b+y_{\rm Ly\alpha} T_{\rm Ly\alpha}}
	{1+y_c+y_{\rm Ly\alpha}},
	\label{eq:tspin}
\end{equation}
where, $y_{\rm Ly\alpha}$ represents the Wouthuysen-Field effect in $T_s$. The quantities $y_c$, and $T_{\rm Ly\alpha}$ are the collisional coupling parameters and the Lyman-$\alpha$ background temperature respectively \cite{BH_21cm_1} 
whereas $y_c$ and $y_{\rm Ly\alpha}$ are defined as 
$y_c=\frac{C_{10}T_{\star}}{A_{10} T_b}$ and 
$y_{\rm Ly\alpha}=\frac{P_{10}T_{\star}}{A_{10} T_{\rm Ly\alpha}}e^{0.3 
	\times (1+z)^{1/2} T_b^{-2/3} \left(1+\frac{0.4}{T_b}\right)^{-1}}$ 
\cite{BH_21cm_2,Yuan_2010,Kuhlen_2006,Yang_2019}. $C_{10}$ is the collision deexcitation 
rate of the hyperfine level and $P_{10}\approx1.3\times 10^{-21}S_{\alpha}J_{-21}\,{\rm s^{-1}}$ is 
the deexcitation rate due to Lyman-$\alpha$, where $S_{\alpha}$ and $J_{-21}$ 
are the spectral distraction factor \cite{salpha} and the Lyman-$\alpha$ 
background intensity respectively. The background intensity $J_{-21}$ is estimated as described in \cite{jalpha}.

It is to be noted that the effects of the Lyman-$\alpha$ photons from the 
first stars, play a significant role for the shape of the spin temperature 
curve when $z \lesssim 25$. 
In the epoch of the cosmic dark age, the CMB photons contribute to flip 
the spin state of hydrogen atoms. As a consequence, the spin temperature ($T_s$) gets closer to the CMB temperature (over inverse of the redshift). However, later ($z \lesssim 25$), the Lyman-$\alpha$ photons from the new-born stars lead to quick transition of the spin temperature $T_s=T_b$. As a result, the spin 
temperature becomes almost equal to the baryon temperature during the age of 
cosmic dawn \cite{tstb2,tstb1} (Fig.~\ref{fig:tspin}). 
This phenomenon is known as the Wouthuysen-Field effect. The strength of the Wouthuysen-Field effect depends on the rate of scattering of the Lyman-$\alpha$ photons in the IGM \cite{lya001}.

\section{Calculations and Results} \label{sec:result} 
\begin{figure}
	\centering
	\includegraphics[width=0.7\columnwidth]{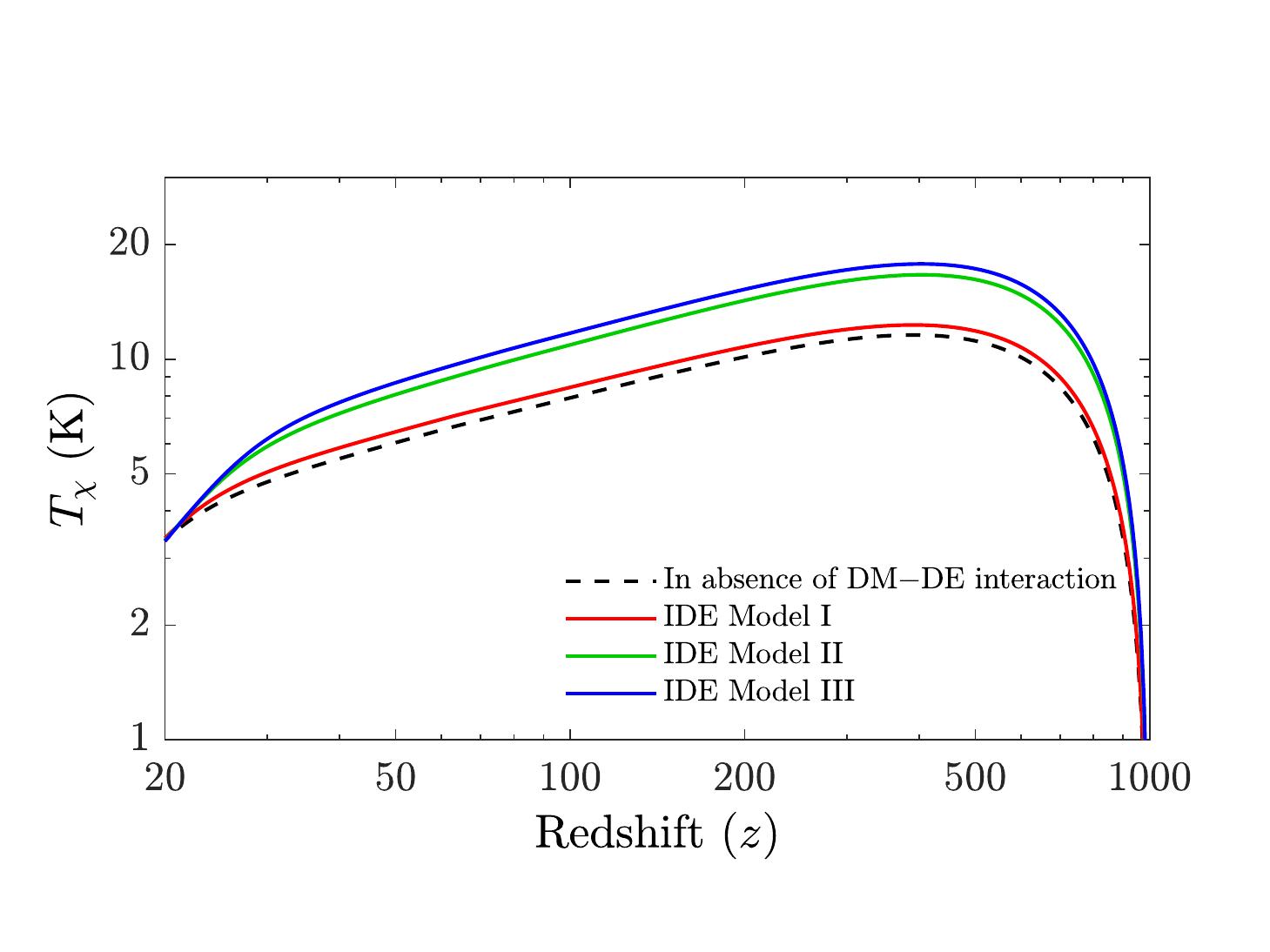}
	\caption{\label{fig:tchi} Variation of dark matter temperature ($T_{\chi}$) with redshift $z$ for different IDE models. For all these plots we consider $\lambda = 0.05$, $\mathcal{M_{\rm BH}}=10^{14}$g, $\beta_{\rm BH} = 10^{-29}$, $m_{\chi}=1$ GeV, $\sigma_{41}=1$, while $\omega$ is taken from Table~\ref{tab:constraints} for different IDE models.}
\end{figure}

In this work, we explore the 21-cm anomalous hydrogen absorption 
line in reionization era by considering the possible simultaneous 
effects of Hawking 
radiation form PBHs along with the DM - baryon interaction. In doing this, we numerically solve seven coupled equations, (Eqs.~\ref{eq:rho_chi}, \ref{eq:rho_de}, \ref{eq:PBH}, \ref{eq:T_chi}, \ref{eq:T_b}, \ref{eq:xe} and \ref{eq:V_chib}) simultaneously. 
The evolutions is initiated at redshift $z=1010$ as described in the work of Mu\~{n}oz \cite{munoz}, when the baryons are assumed to be tightly coupled with the CMB photons (i.e. $T_b \approxeq T_{\chi}$). At $z=1010$, the initial relative velocity is considered to be $\sim$29 km/s and the temperature of DM fluid is assumed to be $T_{\chi}=0$. It is to be mentioned that, even if a slightly warm dark matter candidate is taken into account, the evolution remains almost the same (it has also been discussed by Mu\~{n}oz \cite{munoz}). Since we consider two matter fluids (dark matter fluid and baryon fluid) in this system, the cooler fluid (dark matter fluid) tends to heat up at the expense of the temperature of the comparatively warmer fluid (baryon fluid) as an outcome of the temperature exchange between them (even without IDE effect (Fig.~\ref{fig:tchi})). Although this heating rate is essentially proportional to ($T_b-T_{\chi}$), the heating rate gets perturbed in presence of the drag term ($V_{\chi b}$). A detailed analysis regarding this thermal effect has been addressed in \cite{munoz}.
\begin{figure}
	\centering
	\begin{center}
		\includegraphics[width=0.7\columnwidth]{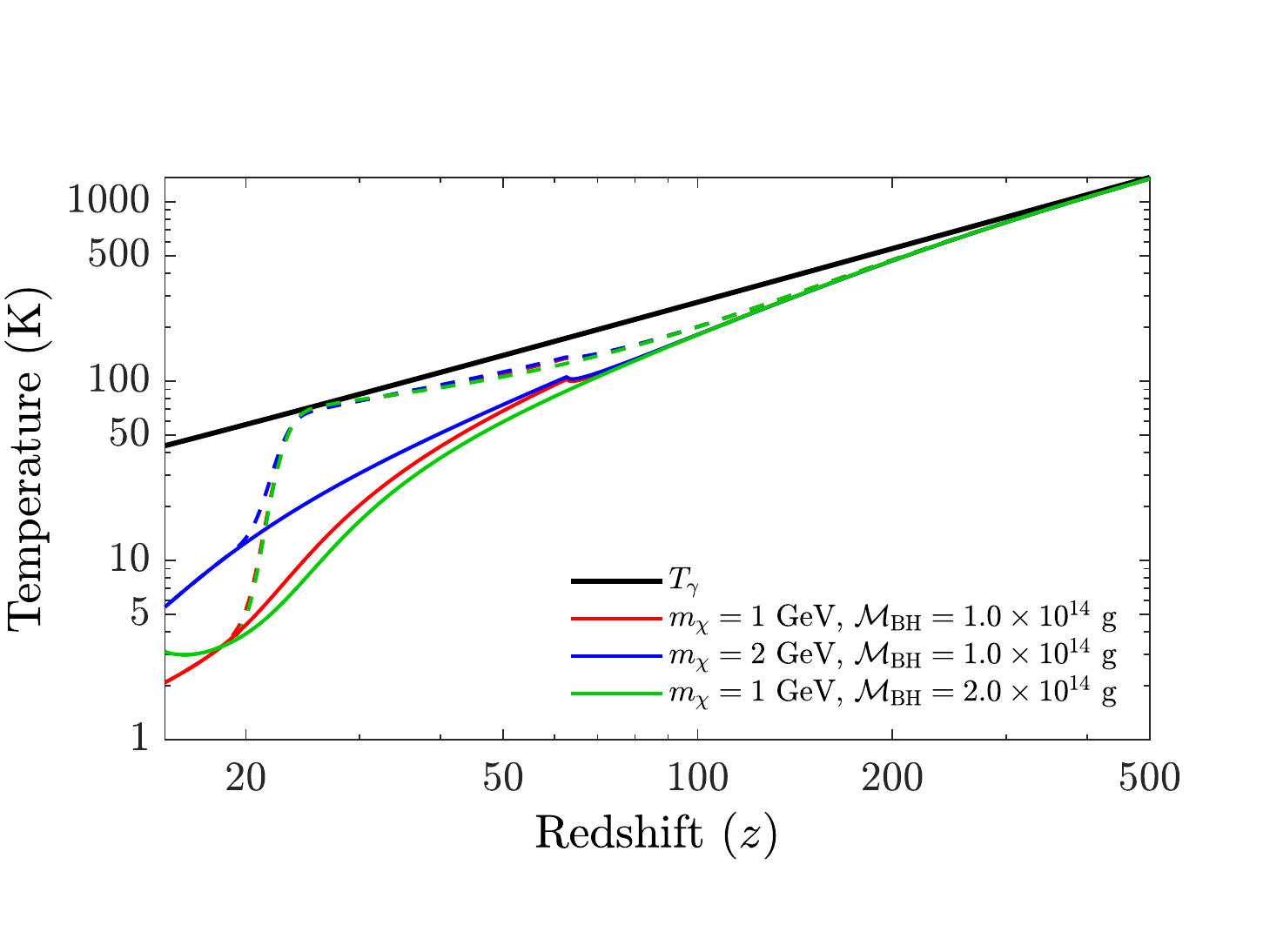}
	\end{center}
	\caption{\label{fig:tspin} The variation of baryon temperature $T_b$, 
	background temperature $T_{\gamma}$, spin temperature $T_s$ with redshift $z$. 
	The black solid line represents the variation of $T_{\gamma}$ and the coloured 
	solid lines and coloured dashed lines correspond to the variations of $T_b$ and $T_s$ respectively with $z$, for different sets of dark matter mass $m_{\chi}$ 
	and PBH mass $\mathcal{M}_{\rm BH}$. Note that for $T_s$, the plots for all 
	three sets almost coincide. For both $T_b$ and $T_s$, the computations 
	are made with Model I (Table~\ref{tab:stability} and \ref{tab:constraints}) 
	only.}
\end{figure}
In Fig.~\ref{fig:tspin}, the evolution of baryon temperature ($T_b$), CMB 
temperature ($T_{\gamma}$) and the corresponding spin temperature ($T_s$) are 
plotted as a function of redshift $z$. The solid red line in Fig.~\ref{fig:tspin} describes the baryon temperature ($T_b$). Spin temperature ($T_s$, 
red dashed line) variations in presence of PBHs of mass 
$\mathcal{M}_{\rm{BH}}=10^{14}$ g 
and dark matter mass $m_{\chi}=1$ GeV, with the IDE Model I (model 
parameters are chosen from Table.~\ref{tab:constraints}) is also shown. 
The blue and green solid and dashed lines are for the same with $m_{\chi}=2$~GeV, $\mathcal{M}_{\rm{BH}}=1.0 \times 10^{14}$ g and $m_{\chi}=1$~GeV, $\mathcal{M}_{\rm{BH}}=2.0\times 10^{14}$ g respectively. In all the cases however Model I for DM - DE interaction is used. It can also be seen that, while the variation of $T_b$ with $z$ differ for different choices of $m_{\chi}$ and $\mathcal{M_{\rm BH}}$ below $z\sim 100$. Such variations for $T_s$ are barely observed except very mildly around $z\sim 20$.

\begin{figure*}
	\centering
	\begin{tabular}{cc}
		\includegraphics[width=0.48\textwidth]{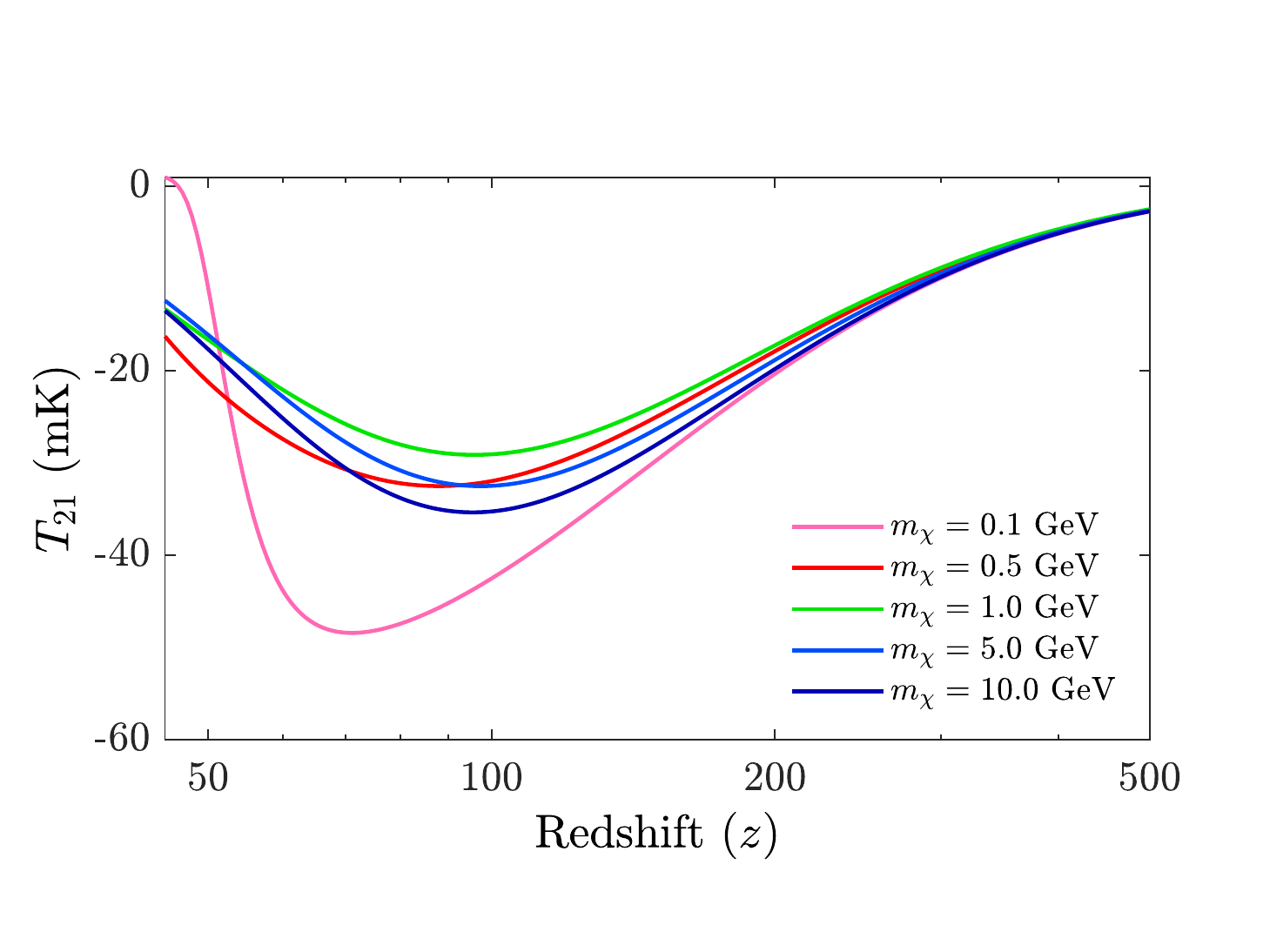}&
		\includegraphics[width=0.48\textwidth]{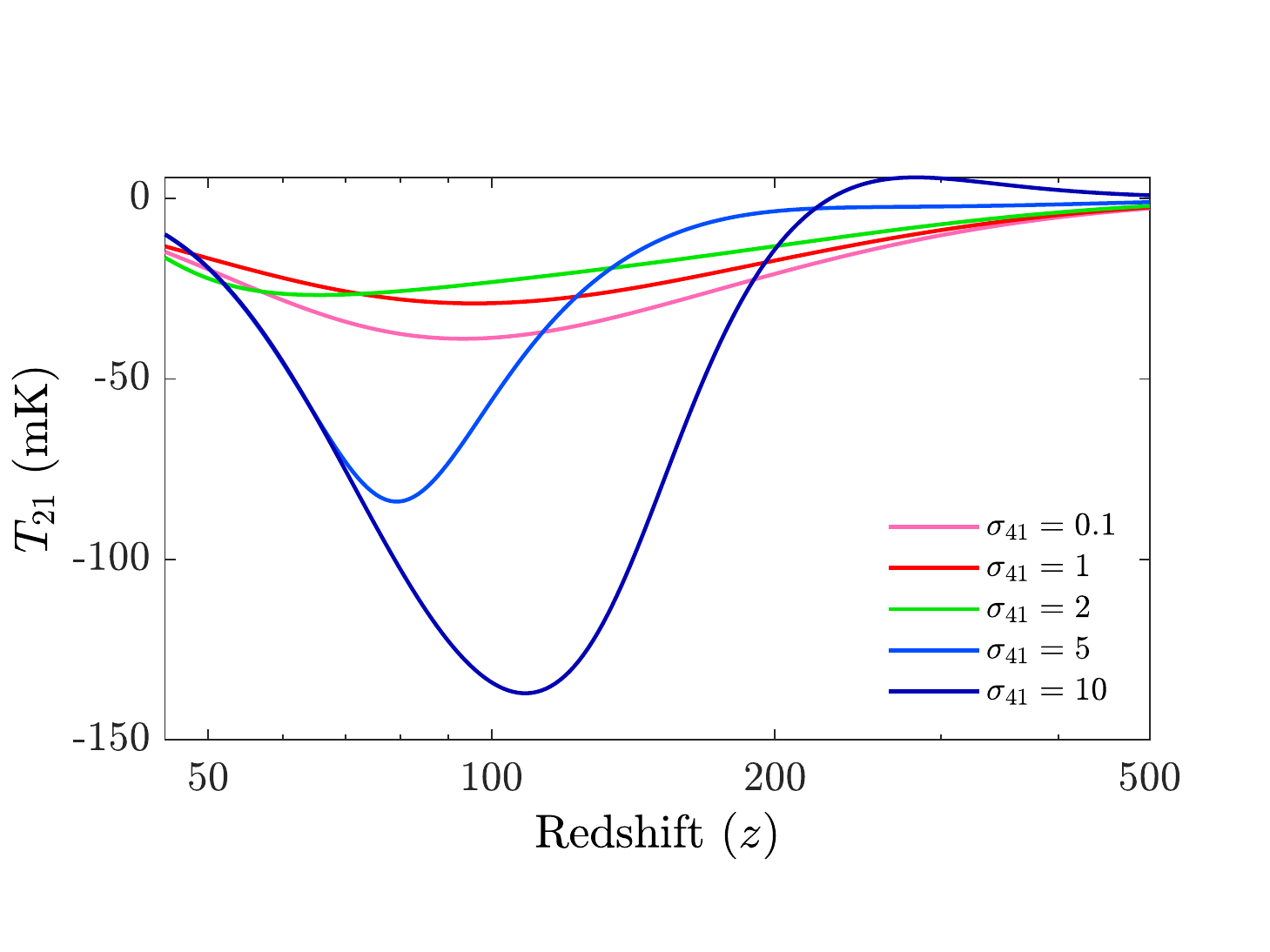}\\
		(a)&(b)\\
	\end{tabular}
	\caption{\label{fig:mchivar} Variations of brightness temperature 
		($T_{21}$) with redshift ($z$) (a) for different values of $m_{\chi}$ 
		(0.1 GeV, 0.5 GeV, 1 GeV, 5 GeV and 10 GeV) in presence of 
PBH mass $\mathcal{M_{\rm BH}}=1.5 \times 10^{14}$ g and $\sigma_{41}=1$. 
Fig.~\ref{fig:mchivar}(b) (right panel) describes the variations of $T_{21}$ 
for different values of $\sigma_{41}$ ($\sigma_{41}=$0.1, 1, 2, 5, 10) 
when the PBH mass $\mathcal{M_{\rm BH}}=10^{14}$ g and dark matter mass 
$m_{\chi}=1$ GeV are chosen.}
\end{figure*}
In Fig.~\ref{fig:mchivar} we demonstrate how the 21-cm brightness temperature 
$T_{21}$ is affected for different dark matter masses $m_{\chi}$ and various 
dark matter - baryon scattering cross-sections 
(different values of $\sigma_{41}$). We plot in Fig.~\ref{fig:mchivar} the variations of $T_{21}$ with $z$ for different values of $m_{\chi}$ for a fixed value of $\sigma_{41}=1$ (Fig.~\ref{fig:mchivar}a) and for different values of $\sigma_{41}$ for a fixed value of $m_{\chi}=1$ GeV (Fig.~\ref{fig:mchivar}b). For both the cases, the PBH mass is fixed at $\mathcal{M_{\rm BH}}=1.5 \times 10^{14}$ g. It is to be noted that, both Fig.~\ref{fig:mchivar}a and Fig.~\ref{fig:mchivar}b show the variation of $T_{21}$ at higher redshift era ($\gtrapprox 50$).

\begin{figure*}
	\centering
	\begin{tabular}{cc}
		\includegraphics[trim=0 40 0 55, clip, width=0.48\textwidth]
		{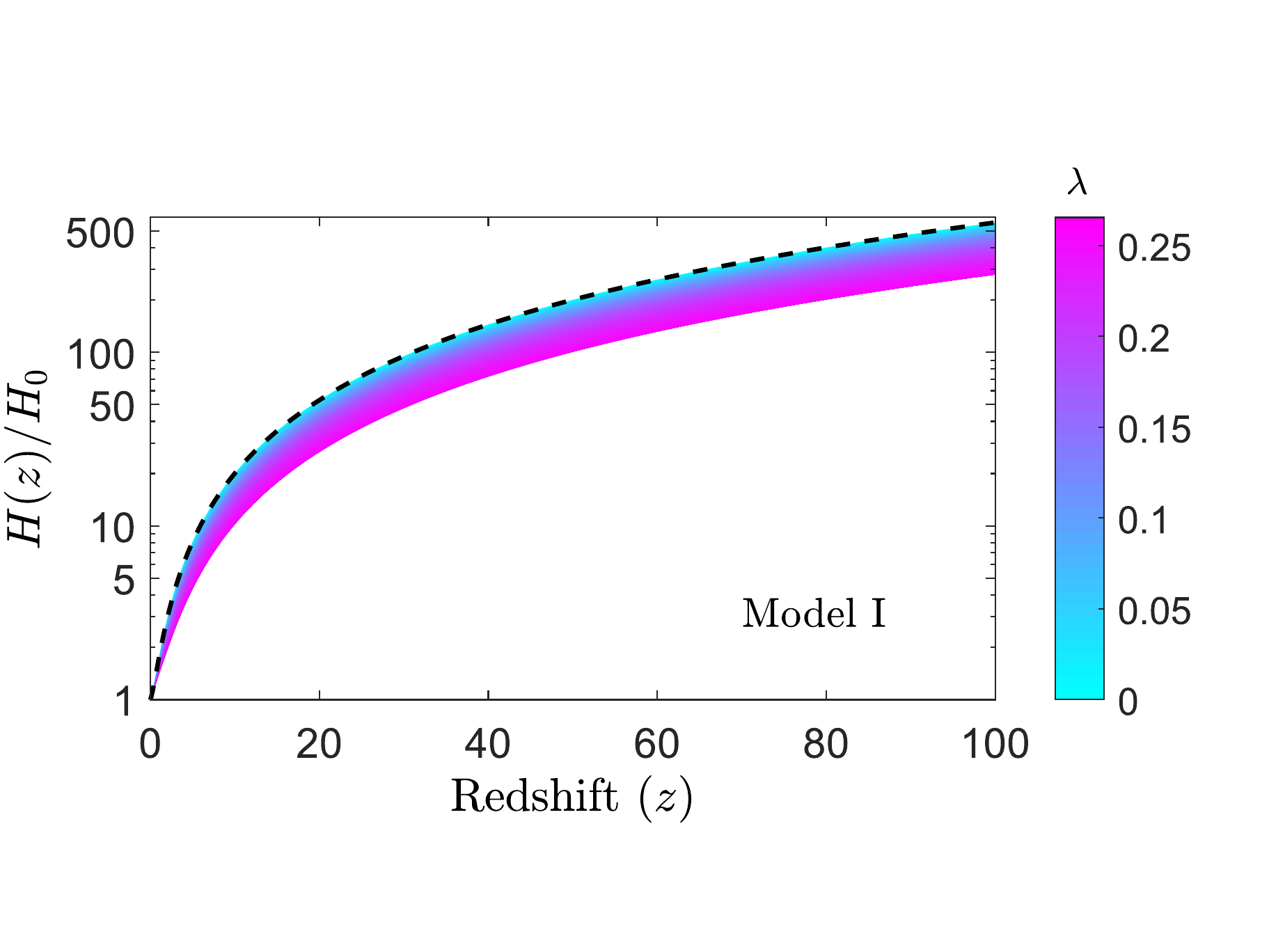}&
		\includegraphics[trim=0 40 10 55, clip, width=0.48\textwidth]
		{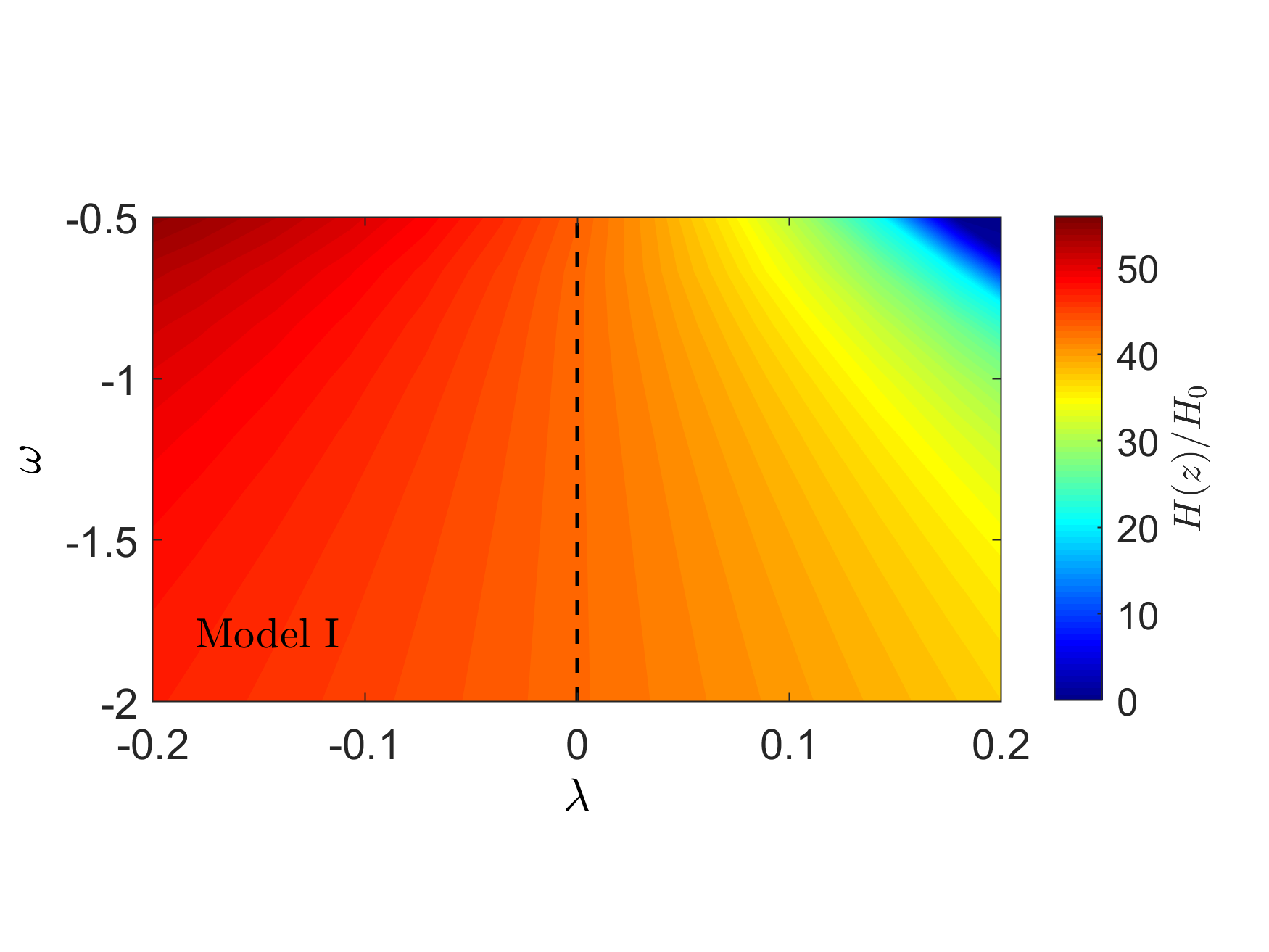}\\
		(a)&(b)\\
		\includegraphics[trim=0 40 0 55, clip, width=0.48\textwidth]
		{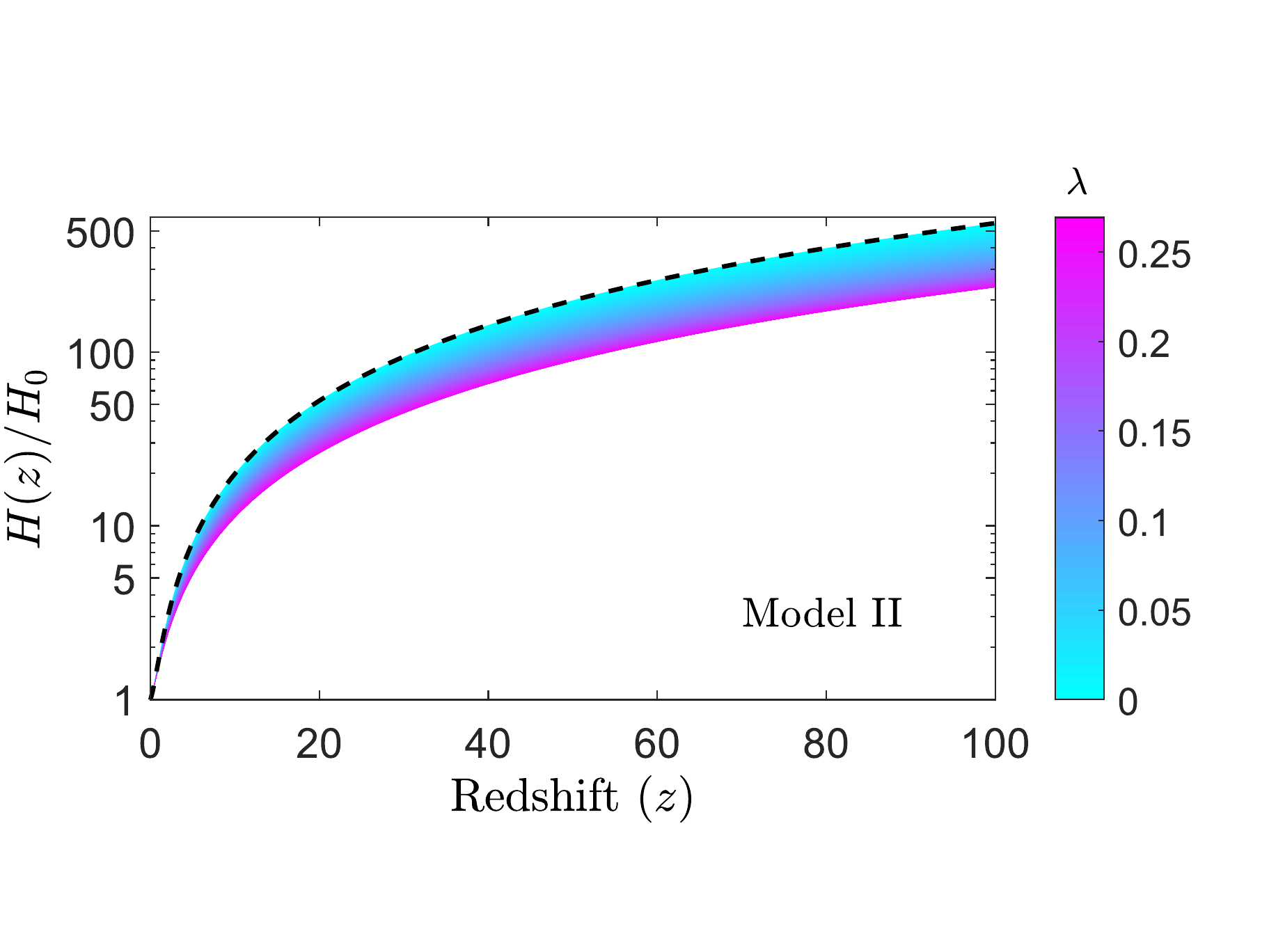}&
		\includegraphics[trim=0 40 10 55, clip, width=0.48\textwidth]
		{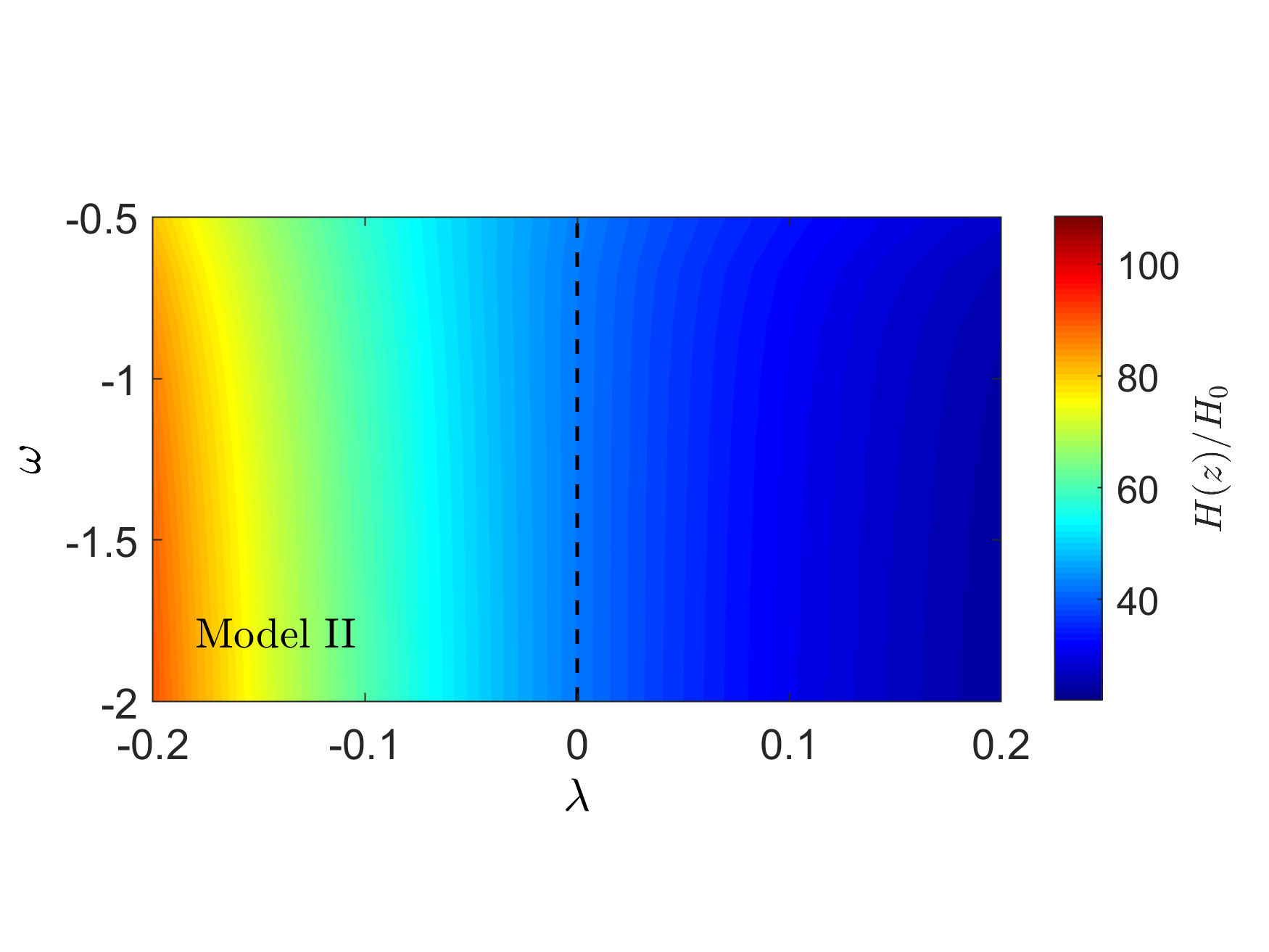}\\
		(c)&(d)\\
		\includegraphics[trim=0 40 0 55, clip, width=0.48\textwidth]
		{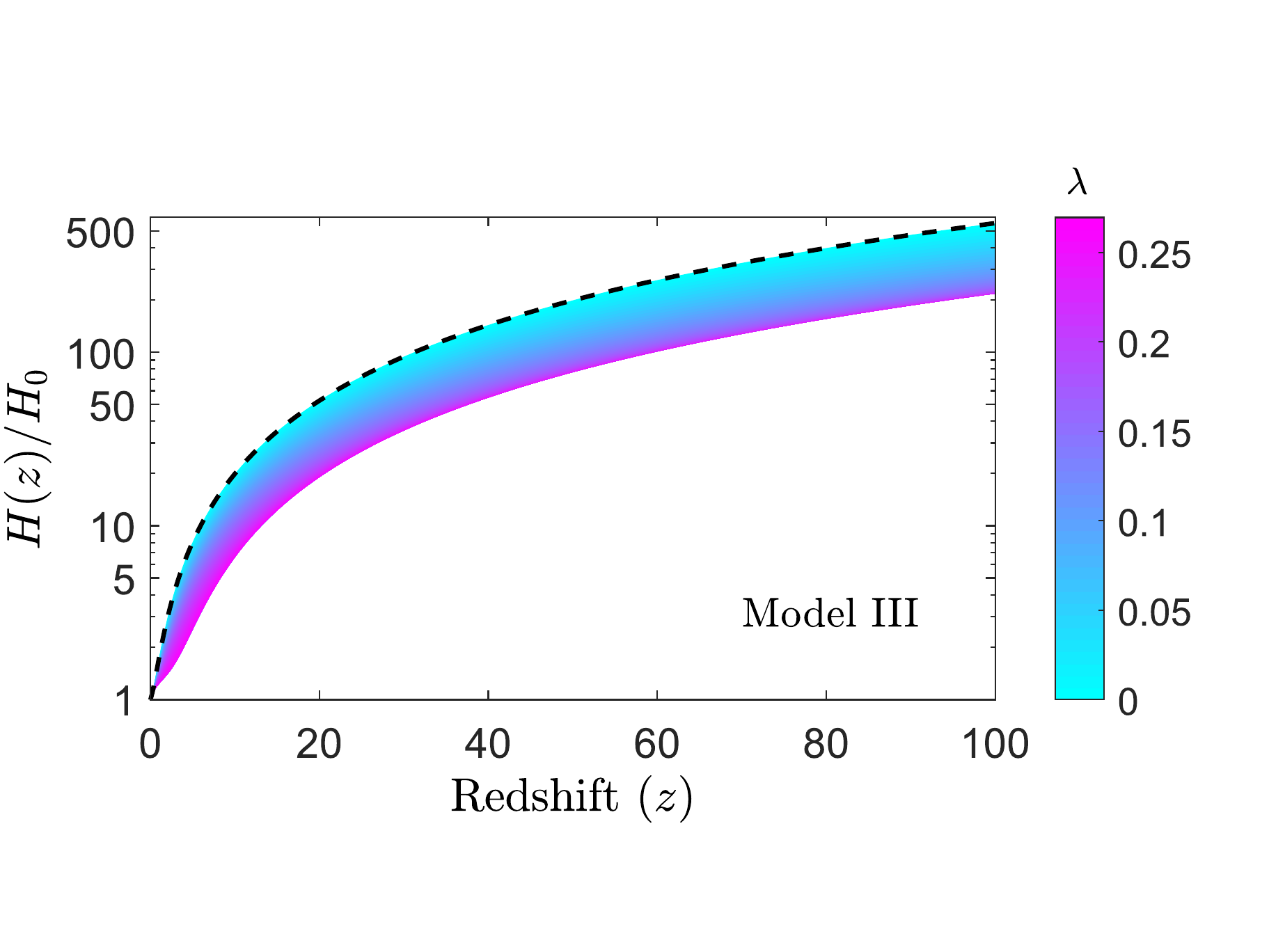}&
		\includegraphics[trim=0 40 10 55, clip, width=0.48\textwidth]{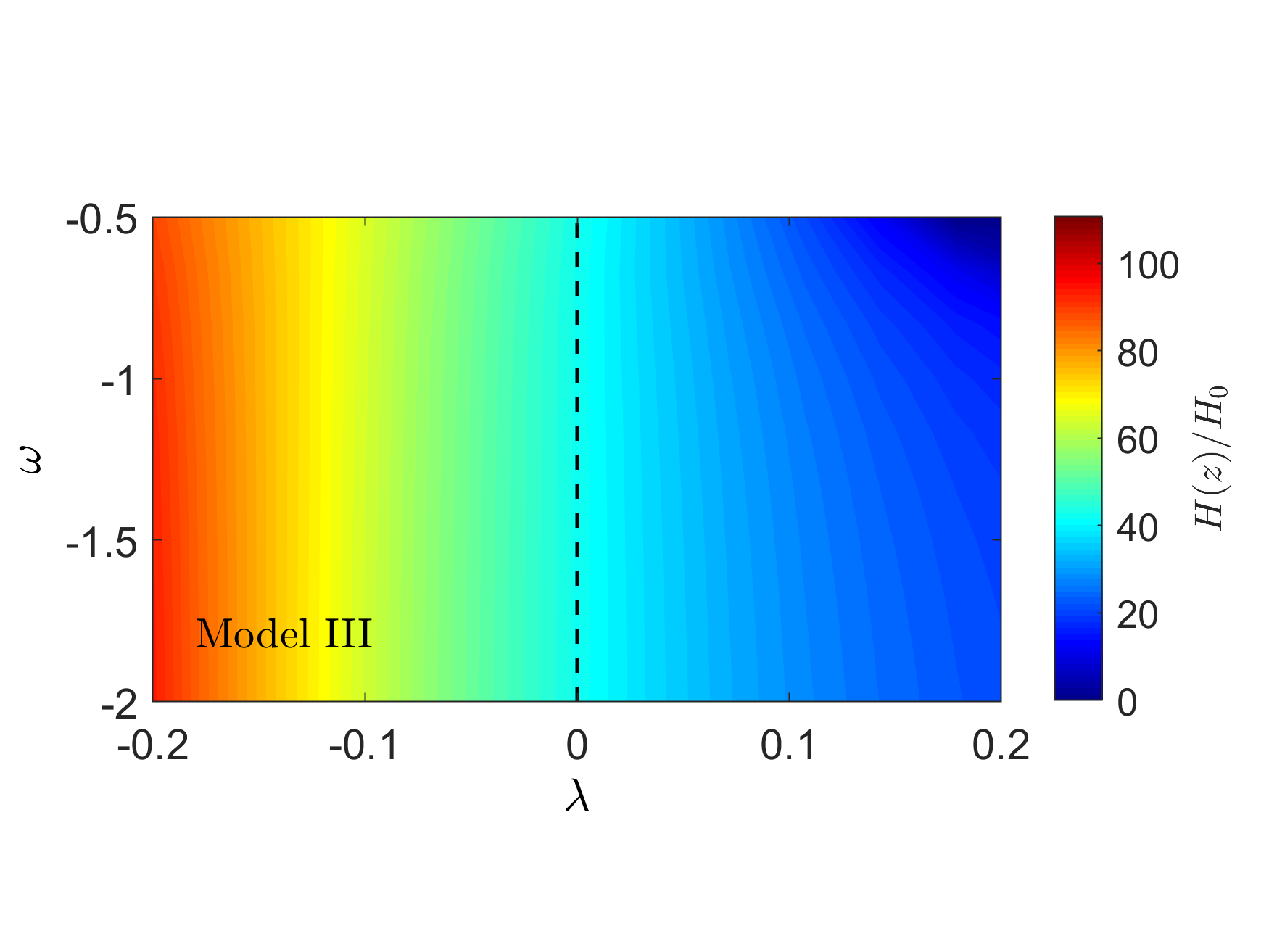}\\
		(e)&(f)\\
	\end{tabular}
	\caption{\label{fig:hubble} Evolution of Hubble parameter ($H(z)$) for 
		different IDE models (Model I, Model II and Model III) 
with different values 
		of coupling parameters $\lambda$. 
The plots on the left panel (Fig.~\ref{fig:hubble}a, 
		\ref{fig:hubble}c and \ref{fig:hubble}e) give the $H(z)$ 
evolution 
		for different values of $\lambda$ where the other parameters ($\omega$ and 
		$H_0$) are chosen from Table~\ref{tab:constraints}. 
The plots on right panel (Fig.~\ref{fig:hubble}b, Fig.~\ref{fig:hubble}d and Fig.~\ref{fig:hubble}f show the variation of $H(z)$ with 
simultaneous variations of the parameters $\lambda$ and 
		$\omega$ (keeping $H_0$ fixed) at redshift $z=17.2$. 
The black dashed line represents the variations of $H(z)$ 
for the $\Lambda$CDM case in each figure.}
\end{figure*}

In presence of the DM - DE interaction, the dark matter and dark energy density parameters  
evolve non-linearly with redshift $z$. As a consequence, remarkable variations in 
the evolution of Hubble parameter $H$ are obtained (see Fig.~\ref{fig:hubble}).
In Fig.~\ref{fig:hubble}, the variation of Hubble parameter as a dimensionless 
quantity $H(z)/H_0$ ($H_0$ is the current value of Hubble parameter) is shown with redshift $z$. The left panel of 
Fig.~\ref{fig:hubble} (Fig.~\ref{fig:hubble}a, \ref{fig:hubble}c and 
\ref{fig:hubble}e) shows  
the variation of Hubble parameter with redshift $z$ for different values of IDE 
coupling parameter $\lambda$, where the different values of $\lambda$ are 
represented by different colours (cyan to magenta, (see colour-bars)). 
On the other hand, 
Figs.~\ref{fig:hubble}b, \ref{fig:hubble}d and \ref{fig:hubble}f 
(right panel of Fig.~\ref{fig:hubble}) show the variations 
of $H(z)/H_0$ with $\lambda$ and the equation of state parameter ($\omega$) 
simultaneously. The plots Fig.~\ref{fig:hubble}a, \ref{fig:hubble}c and 
\ref{fig:hubble}e (left panel of Fig.~\ref{fig:hubble}) correspond to 
Model I, II and III (Tables~\ref{tab:stability}, \ref{tab:constraints}) 
respectively of DM - DE interaction. In the right panel of 
Fig.~\ref{fig:hubble}, the plots of Fig.~\ref{fig:hubble}b, \ref{fig:hubble}d and \ref{fig:hubble}f also correspond to Model I, II and III respectively. In all the three plots of the left panel of Fig.~\ref{fig:hubble} (for Models I, II and III), the values of the DE equation of state parameter $\omega$ for the 
corresponding models are adopted from Table~\ref{tab:constraints}. 
In each of the plots of Fig.~\ref{fig:hubble}, 
the dashed black line represents the $\Lambda$CDM case ($\lambda = 0$). 
Analysing all the 
plots in Fig.~\ref{fig:hubble}, one can conclude that, the evolution of 
Hubble parameter is almost identical in the case of 
Model II and Model III where the variations mostly depend on $\lambda$ (see 
Fig.~\ref{fig:hubble}d and \ref{fig:hubble}e, the dependence on $\omega$ is minimal) 
and the variation is almost linear in nature (see Fig.~\ref{fig:hubble}c and 
\ref{fig:hubble}d). Nonetheless, very small difference can be observed 
between Model II and Model III in Fig.~\ref{fig:hubble}d and \ref{fig:hubble}e 
(also comparing Fig.~\ref{fig:hubble}c and \ref{fig:hubble}d). In contrast, in 
case of Model I, both the parameters, $\lambda$ and $\omega$, 
are equally significant 
in the Hubble parameter evolution (see Fig.~\ref{fig:hubble}b). From Fig.~\ref{fig:hubble}a, it can 
be observed that, initially $H(z)/H_0$ decreases gradually with increasing 
coupling parameter $\lambda$. But later (for higher values of $\lambda$) 
$H(z)/H_0$ suffers rapid fall with increasing $\lambda$. Therefore, although the 
allowed range of $\lambda$ for Model I is $\lambda<-2\omega \Omega_{\chi}$ (see 
Table~\ref{tab:stability}), we limit this range to $0<\lambda<0.25$ for the current work.

\begin{figure}
	\centering
	\includegraphics[width=0.7\columnwidth]{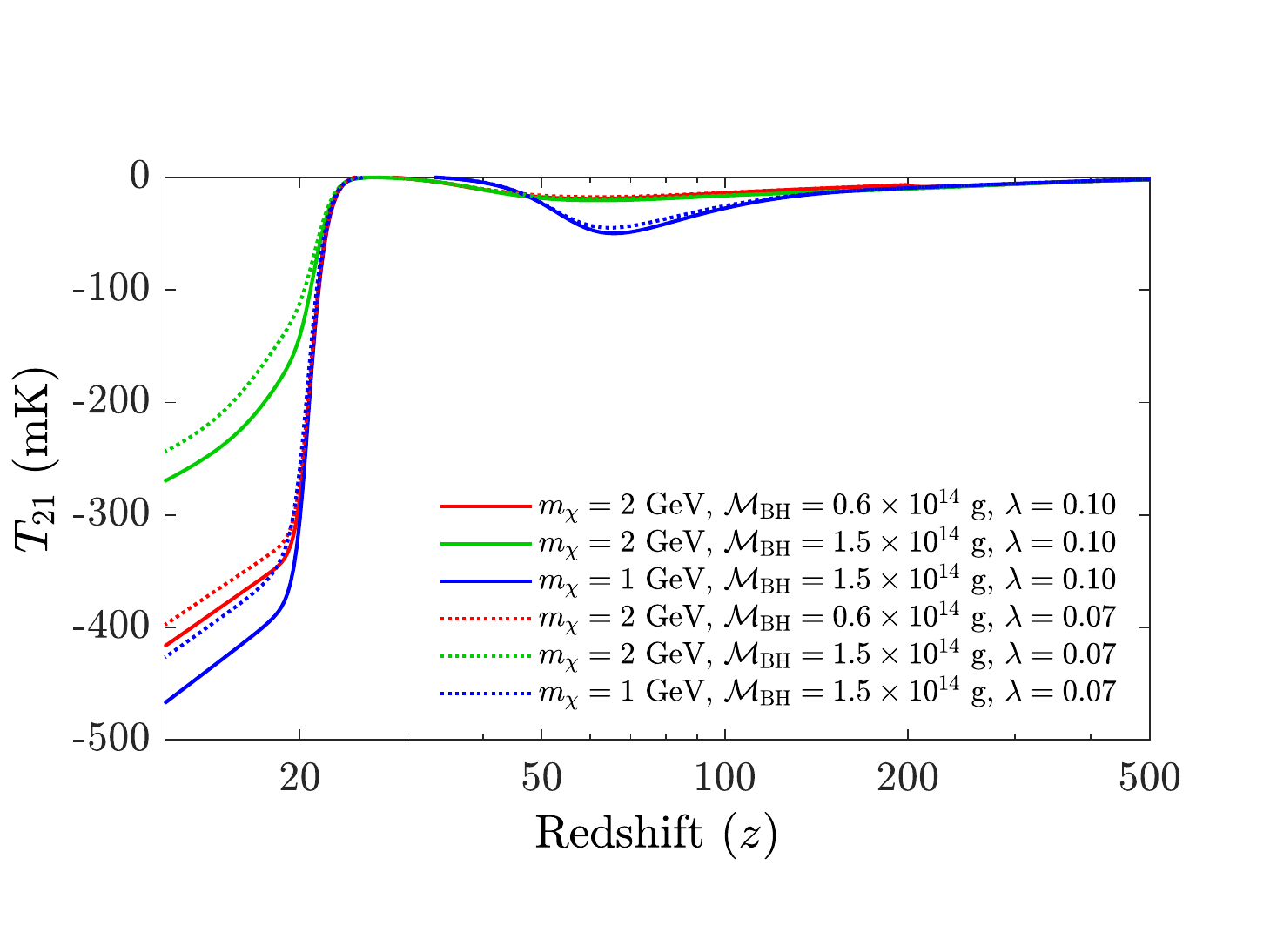}
	\caption{\label{fig:t21} Brightness temperature ($T_{21}$) vs redshift 
		($z$) graph for different values of DM mass ($m_{\chi}$), PBH mass 
		($\mathcal{M_{\rm BH}}$) and  IDE coupling parameter $\lambda$ for 
		Model I.}
\end{figure}

\begin{figure*}
	\centering
	\begin{tabular}{cc}
		\includegraphics[width=0.48\linewidth]{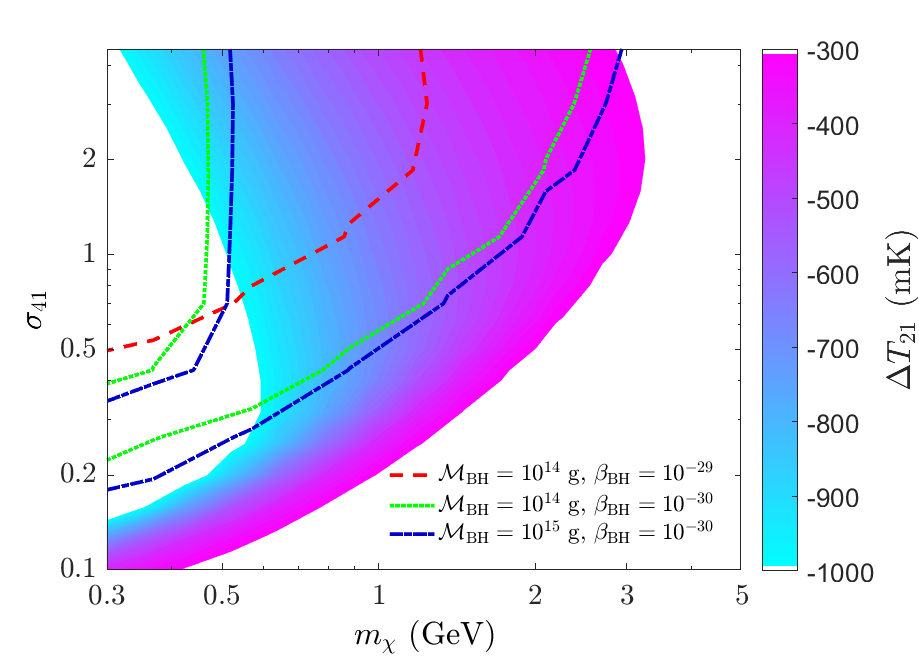}&
		\includegraphics[width=0.48\linewidth]{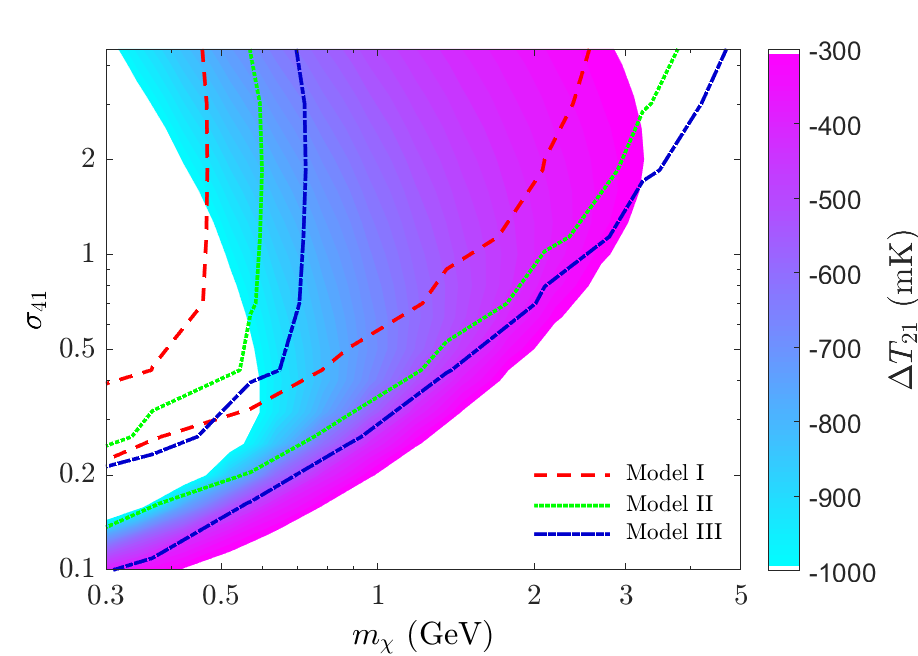}\\
		(a)&(b)\\
	\end{tabular}
	\caption{\label{fig:mchi_sigma} The allowed region of the 21-cm brightness temperature in $m_{\chi}$ - $\sigma_{41}$ space. The coloured shaded 
region represents the allowed region for the case of $\Lambda$CDM model in both the cases. 
In this coloured region, DM - DE interaction has not been considered. 
The coloured lines represent the similar bounds for different values of 
PBH parameters in Fig.~\ref{fig:mchi_sigma}a, for the case of Model I whereas, 
in Fig.~\ref{fig:mchi_sigma}b, the estimated bounds are shown for different 
IDE models (with $\mathcal{M_{\rm BH}}=10^{14}$g, $\beta_{\rm BH}=10^{-30}$ and $\lambda=0.1$)}
\end{figure*}

Incorporating the above mentioned modifications in the density 
parameters and hence the Hubble parameter, the temperature evolution of the Universe is addressed in presence of the PBH and the dark matter - baryon interaction (scattering). The effect of DM-DE interaction is manifested in the evolution of dark matter temperature ($T_{\chi}$). In Fig.~\ref{fig:tchi}, the variation of $T_{\chi}$ with $z$ is shown for different IDE models and compared with the same in absence of the interaction between two dark sector components of the Universe (i.e. dark matter and dark energy). All the plots in Fig.~\ref{fig:tchi} are computed for $\lambda = 0.05$, $\mathcal{M_{\rm BH}}=10^{14}$g, $\beta_{\rm BH} = 10^{-29}$, $m_{\chi}=1$ GeV, $\sigma_{41}=1$, the values of $\omega$ for different IDE models are taken from Table~\ref{tab:constraints}. From Fig.~\ref{fig:tchi} it can be seen that, $T_{\chi}$ increases as a outcomes of the interactions between dark matter and dark energy. This phenomenon indicates that, a significant amount of energy transfers from dark energy to the dark matter, due to DM-DE interaction. It is also noticed that the amount of energy transfer is remarkably high for the case of IDE Model II and IDE model III in comparison to the Model I.
In Fig.~\ref{fig:t21}, we plot the 
21-cm brightness temperatures ($T_{21}$) 
for different values of IDE coupling parameter ($\lambda=$0.07 and 0.10) for dark matter 
masses $m_{\chi} = 2$ GeV and 1 GeV and PBH masses 
$\mathcal{M}_{\rm BH} = 0.6 \times 10^{14}$ g and $0.6 \times 10^{14}$ g while keeping 
$\beta_{\rm BH}$ and $\sigma_{41}$ fixed at $10^{-29}$ and 3 respectively in the 
case of IDE Model I. From this figure (Fig.~\ref{fig:t21}), it can be noticed that, 
besides DM mass and PBH mass, IDE coupling parameter $\lambda$ also modifies 
the brightness temperature remarkably. A detail study for the variation due 
to $\sigma_{41}$ has been carried out in Fig.~\ref{fig:mchi_sigma}. 

The allowed ranges of the dark matter mass $m_{\chi}$ and 
$\sigma_{41}$, that satisfy the EDGES result for $T_{21}$ are also estimated. 
The allowed regions in $m_\chi - \sigma_{41}$ parameter space are shown in 
Fig.~\ref{fig:mchi_sigma} for different PBH masses (Fig.~\ref{fig:mchi_sigma}a) and also for three 
different IDE models (Fig.~\ref{fig:mchi_sigma}b). The contour plots are generated 
for a fixed value of $z=17.2$. In what follows the brightness temperature 
at $z=17.2$ is represented by $\Delta T_{21}$ unless otherwise mentioned.
The EDGES limit of the brightness temperature ($T_{21}$) at  
$z \sim 17.2$ lie within the range of  $-300$ mK $\geq T_{21} \geq -1000$ mK. 
In the current analysis, we estimate the limit for $m_{\chi}$ and $\sigma_{41}$ 
using the above mentioned constraint ($-300$ mK $\geq T_{21} \geq -1000$ mK). 
In Fig.~\ref{fig:mchi_sigma}a, the calculated 
allowed zones in the 
$m_{\chi}$ - $\sigma_{41}$ plane are plotted for $\Lambda$CDM model 
using the colour code 
(coloured shaded region), where the individual colours within the coloured 
shaded region indicate the different values of 
$\Delta T_{21}$ (colour code is described in the 
corresponding colour-bar). This may be mentioned that the limits on 
$m_{\chi}$ obtained from these contour plots agree with a similar 
calculation given in \cite{rennan_3GeV} (i.e. $m_{\chi} \leq 3$ GeV). 
In both the plots (a) and (b) of Fig.~\ref{fig:mchi_sigma}, the coloured 
contour plots are generated by varying $m_{\chi}$ and $\sigma_{41}$ and 
computing $\Delta T_{21}$. 
For the coloured shaded regions (not the line contours) in both 
Fig.~\ref{fig:mchi_sigma}a and \ref{fig:mchi_sigma}b, only the dark matter - baryon interaction 
is considered and the relevant coupled differential parameters are 
simultaneously solved numerically. The other contour plots generated using all 
the effects considered here namely, the DM - baryon interaction, 
DM - DE interaction and PBH evaporations are shown by line contour plots 
(allowed region enclosed by lines) in both Fig.~\ref{fig:mchi_sigma}a and Fig.~\ref{fig:mchi_sigma}b. 
In these latter cases the allowed regions are bound by different pairs of lines in 
$m_{\chi}$ - $\sigma_{41}$ plane. In Fig.~\ref{fig:mchi_sigma}a, three different line 
contours are for three different sets of values of PBH masses and 
$\beta_{\rm BH}$ when Model I (Table~\ref{tab:stability} and \ref{tab:constraints}) is used for DM - DE interaction 
in all the three line contour plots. In Fig.~\ref{fig:mchi_sigma}b, the three allowed contours 
(area enclosed by different lines) in $m_\chi - \sigma_{41}$ plane are 
generated with three IDE models (Model I, II and III, Table~\ref{tab:stability}, \ref{tab:constraints}) while 
the PBH mass and $\beta_{\rm BH}$ are kept fixed at values of $10^{14}$ g 
and $10^{-29}$ respectively. 
From Fig.~\ref{fig:mchi_sigma}a it can be seen that the region enclosed 
by red dashed line corresponds to the $m_{\chi}$ - $\sigma_{41}$ allowed region when PBH mass $\mathcal{M_{\rm BH}} = 10^{14}$ g and 
$\beta_{\rm BH}=10^{-29}$ are chosen with Model I for DM - DE interaction. 
Similarly, the region enclosed by the green dotted lines specifies 
the allowed region when $\mathcal{M_{\rm BH}}=10^{15}$ g, 
$\beta_{\rm BH}=10^{-30}$ are considered. From Fig.~\ref{fig:mchi_sigma}a this can be observed that as 
$\mathcal{M_{\rm BH}}$ decreases (Hawking radiation increases) the allowed 
region shifts towards higher $\sigma_{41}$ and lower $m_\chi$ region. 
A similar trend is also observed for the IDE Model I (Fig.~\ref{fig:mchi_sigma}b). On the other hand, in the case of IDE Model II and III, the allowed region shifts toward higher values of $m_{\chi}$.

\begin{figure*}
	\centering
	\begin{tabular}{ccc}
		\includegraphics[trim=0 20 80 30, clip, height=0.23\textwidth]{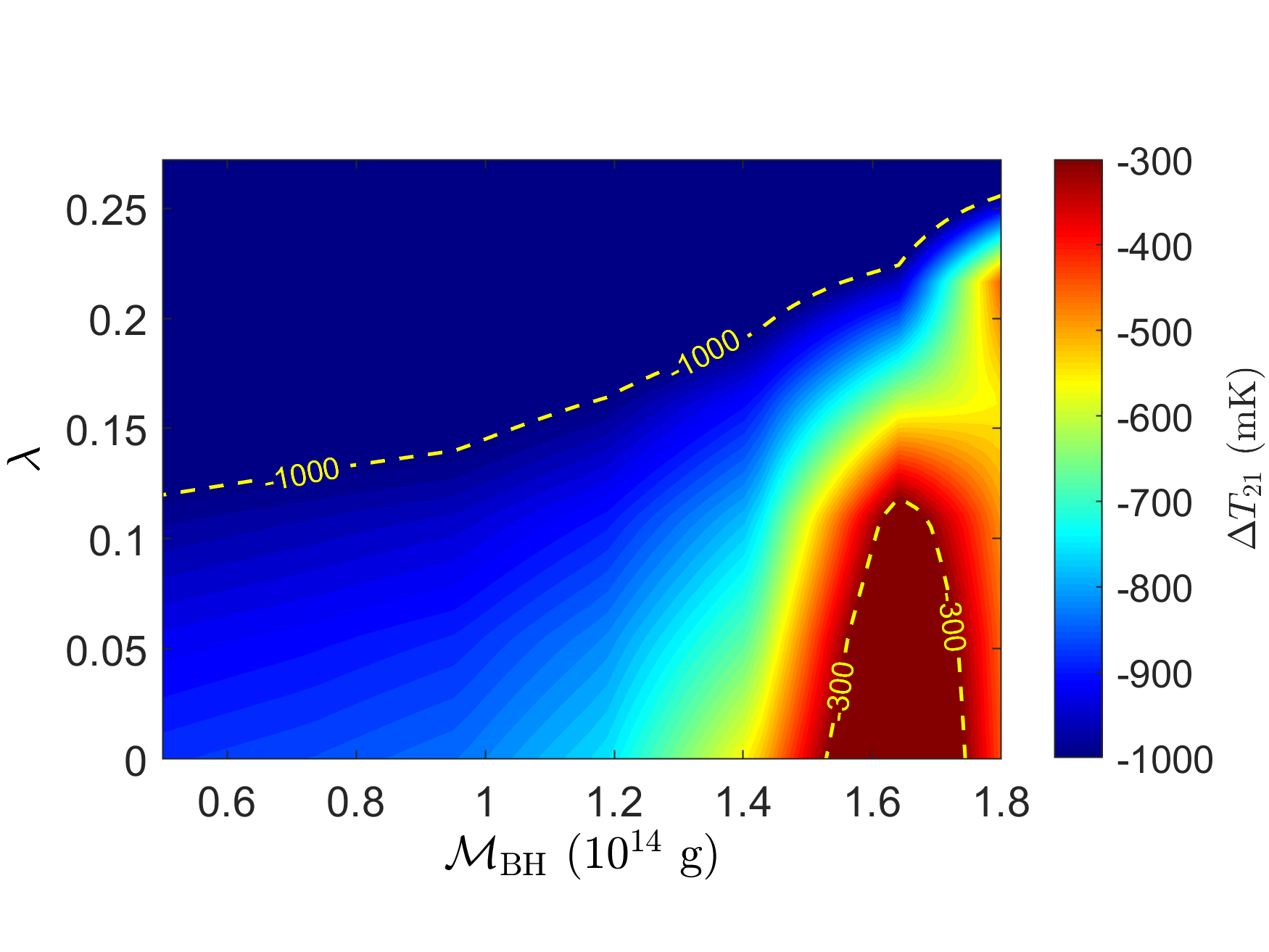}&
		\includegraphics[trim=0 20 80 30, clip, height=0.23\textwidth]{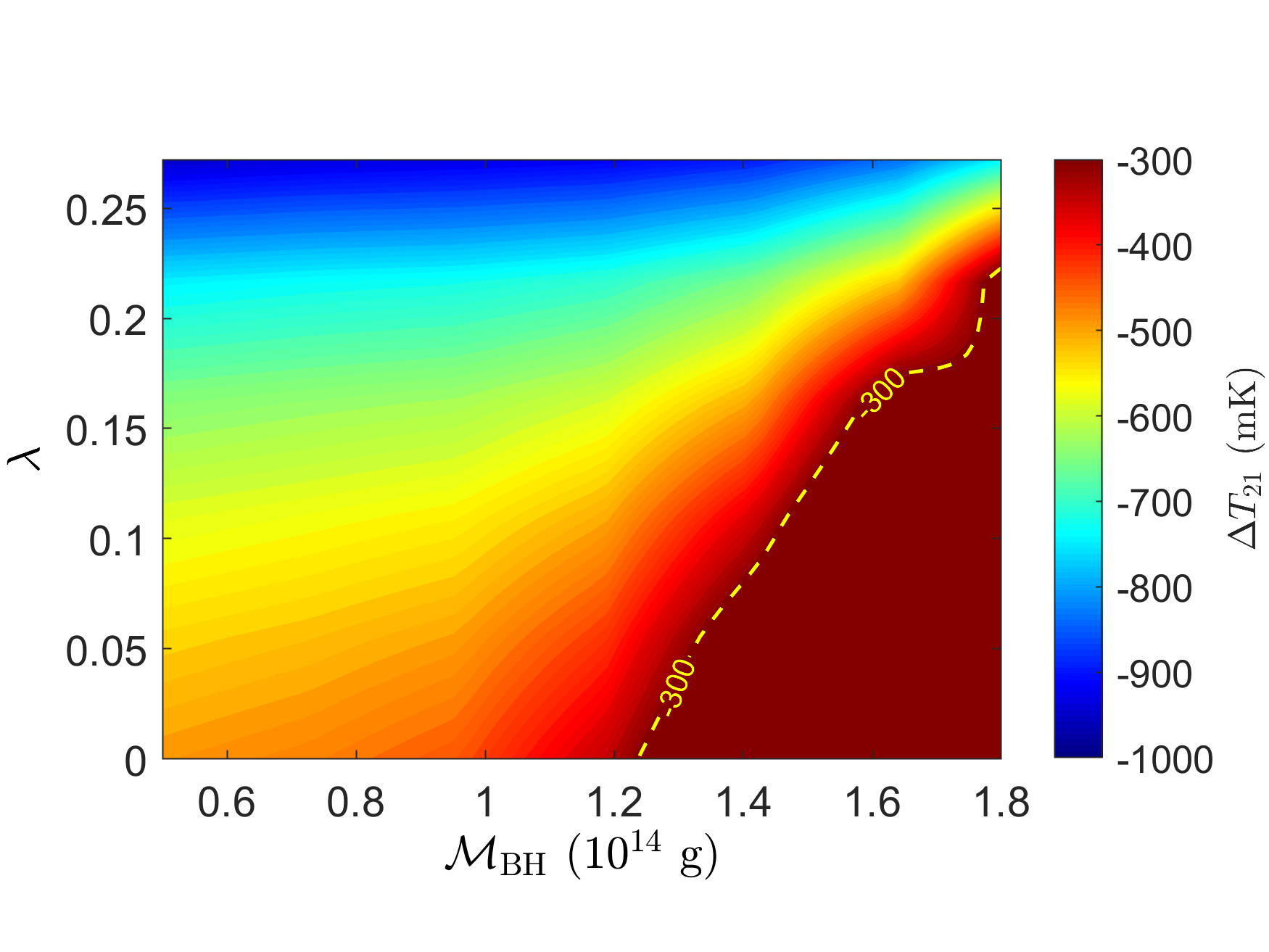}&
		\includegraphics[trim=0 20 0 30, clip, height=0.23\textwidth]{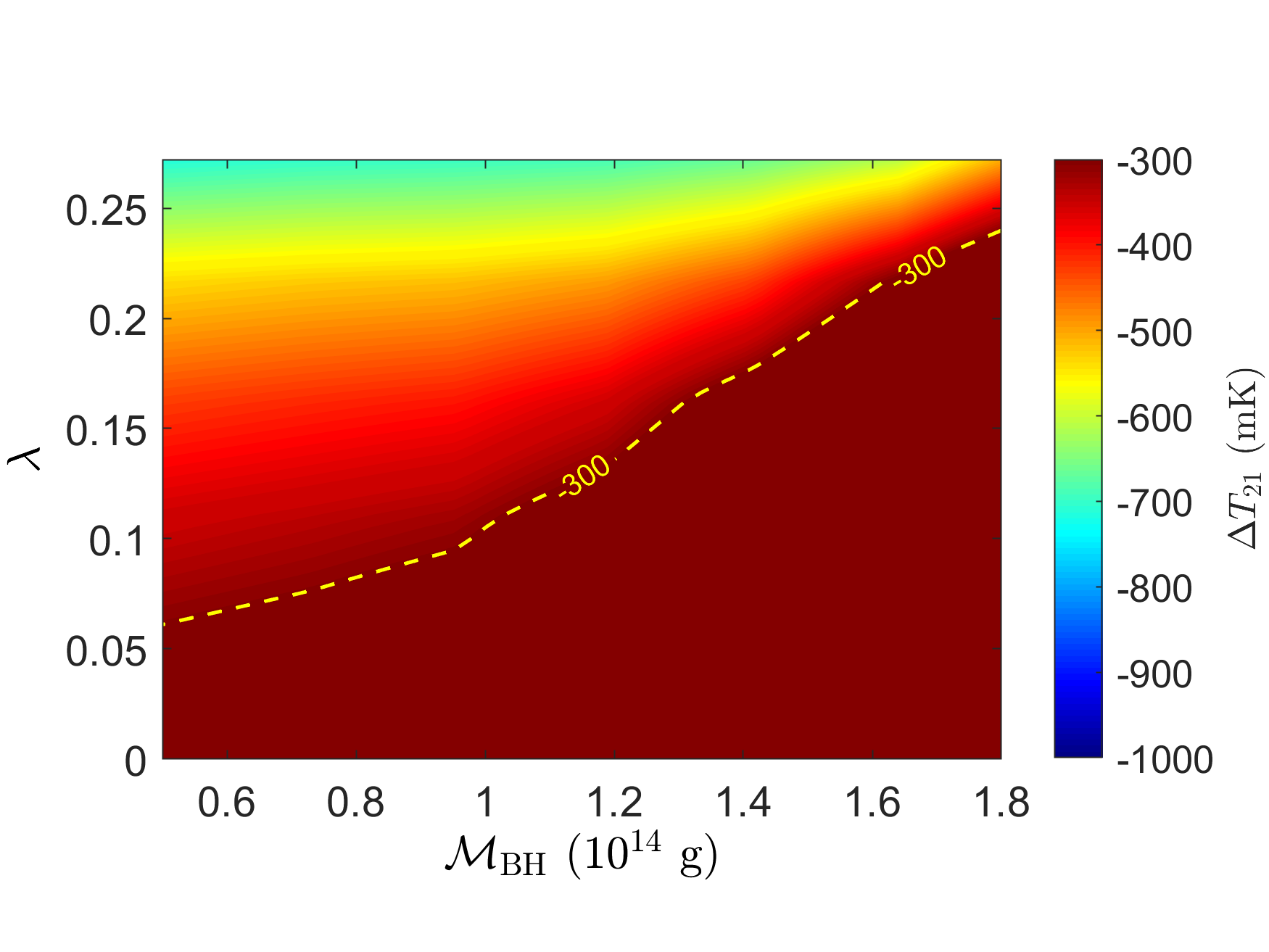}\\
		(a)&(b)&(c)\\
		\includegraphics[trim=0 20 80 30, clip, height=0.23\textwidth]{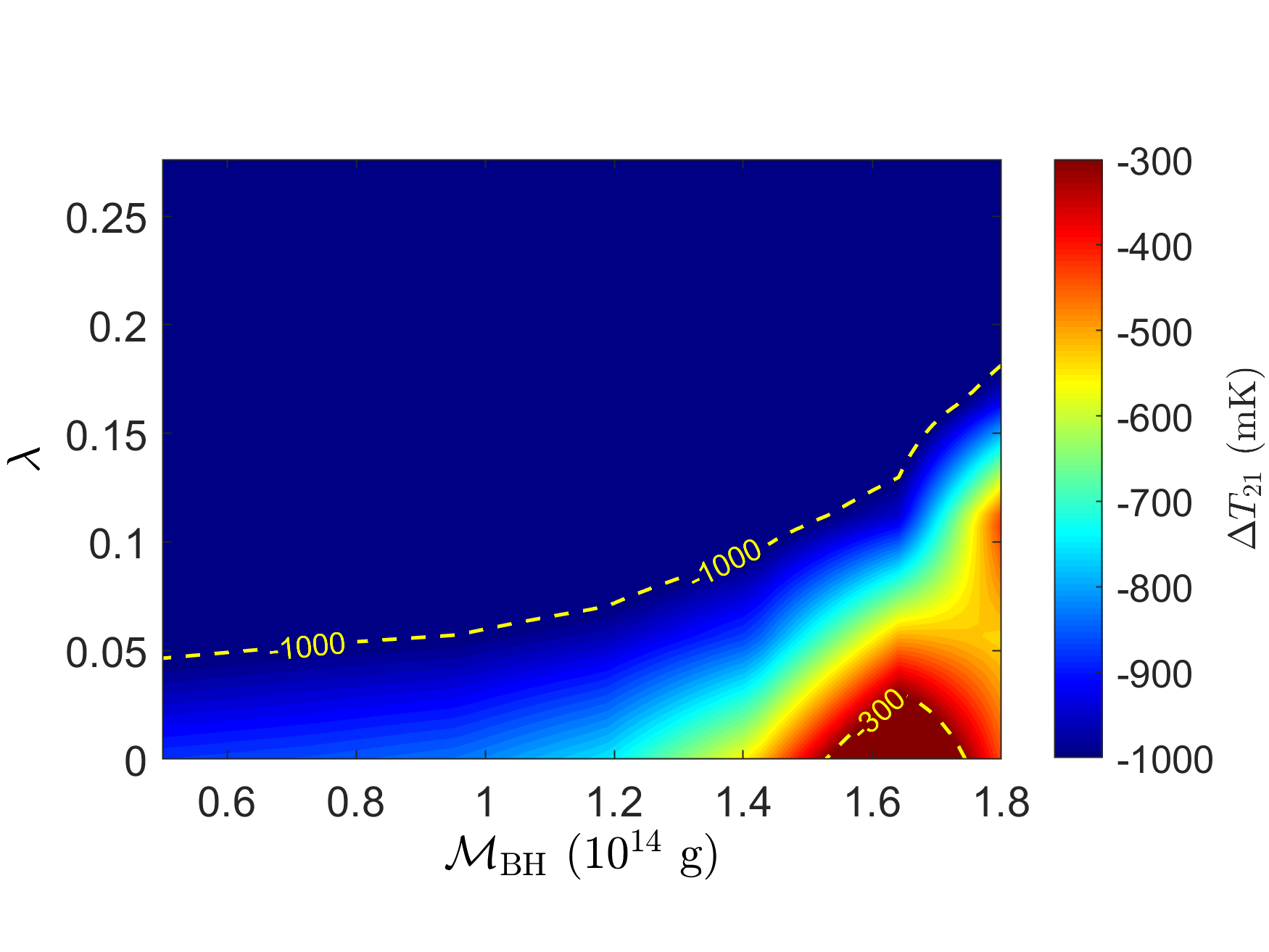}&
		\includegraphics[trim=0 20 80 30, clip, height=0.23\textwidth]{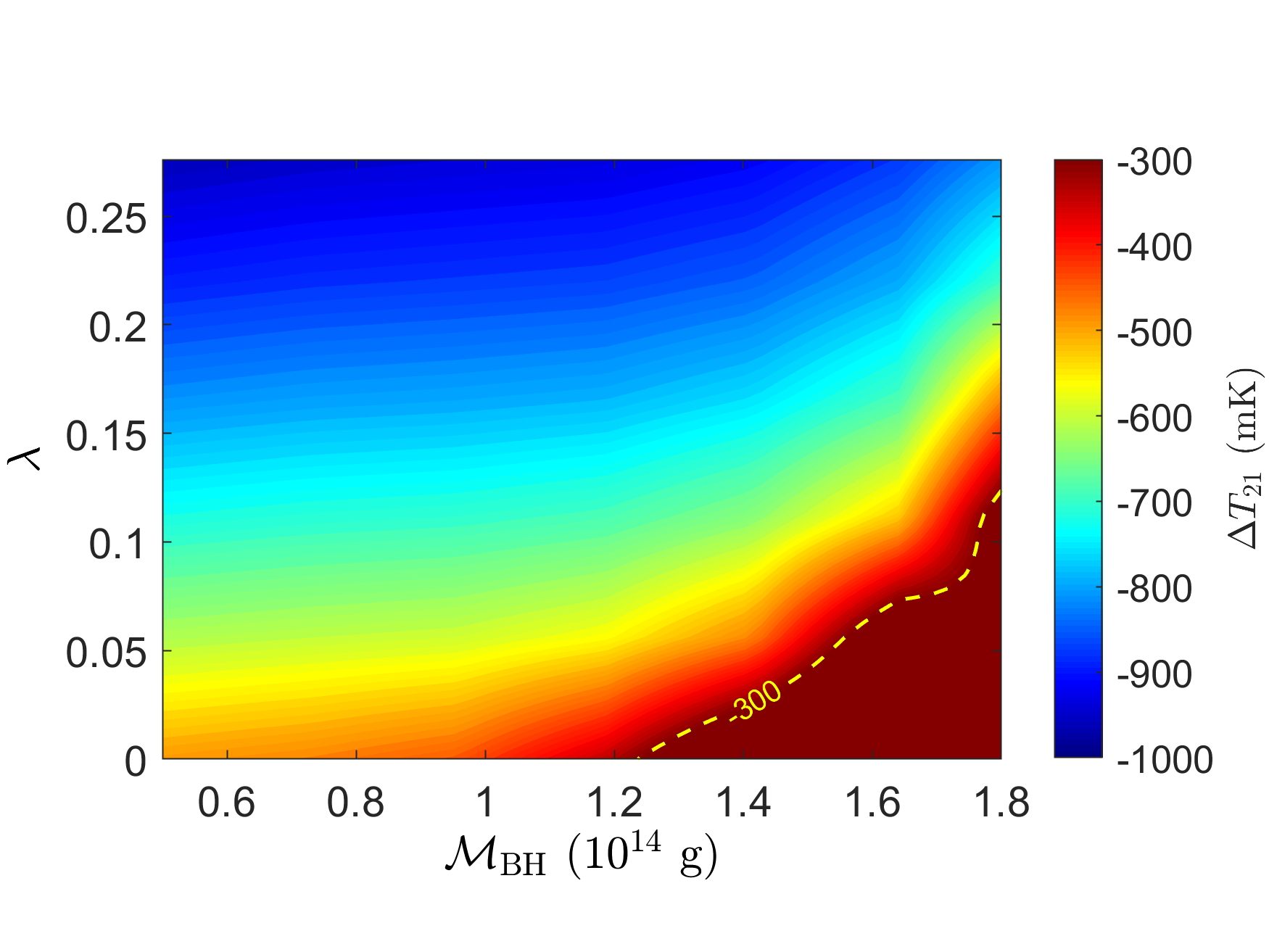}&
		\includegraphics[trim=0 20 0 30, clip, height=0.23\textwidth]{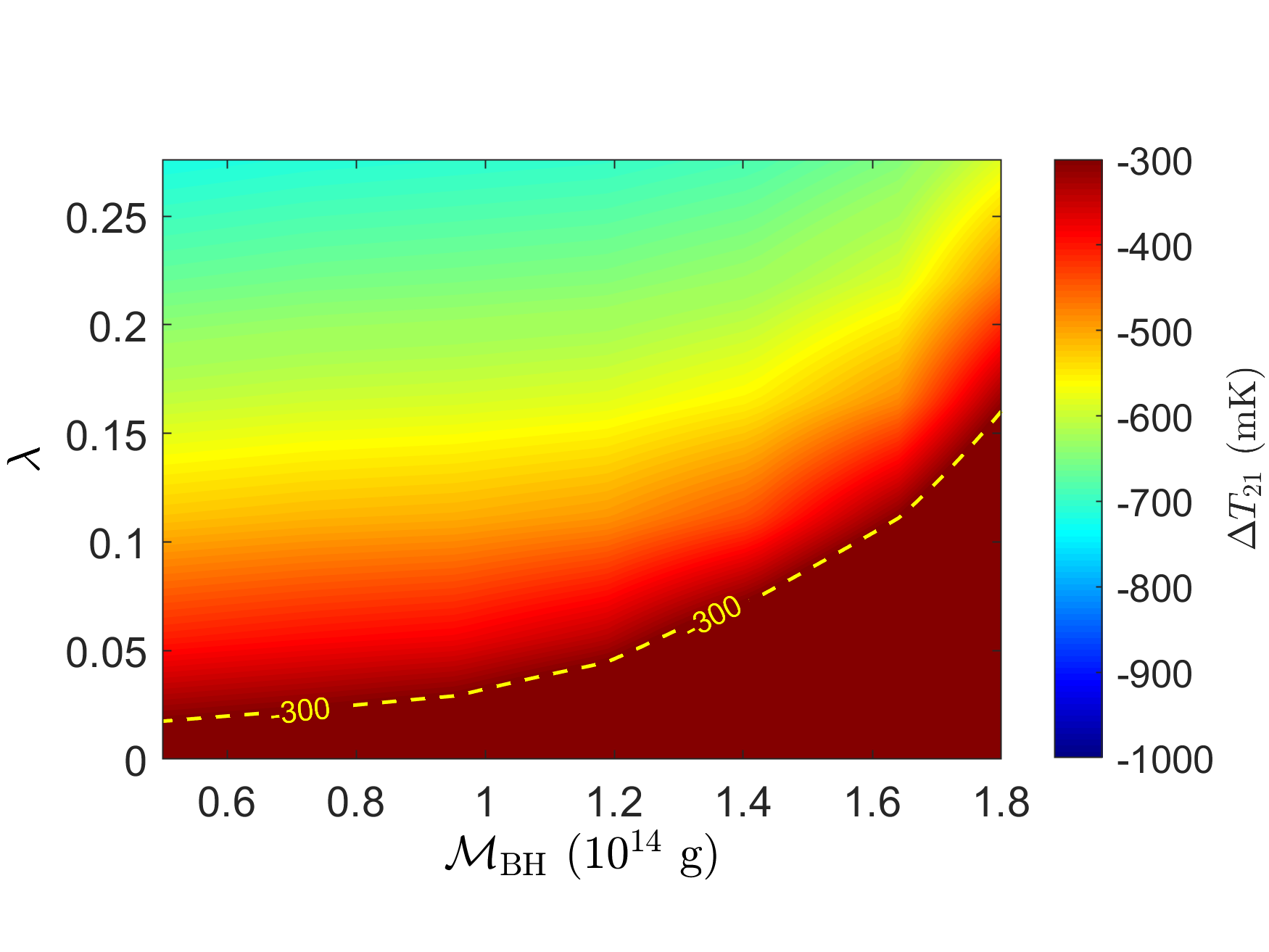}\\
		(d)&(e)&(f)\\
		\includegraphics[trim=0 20 80 30, clip, height=0.23\textwidth]{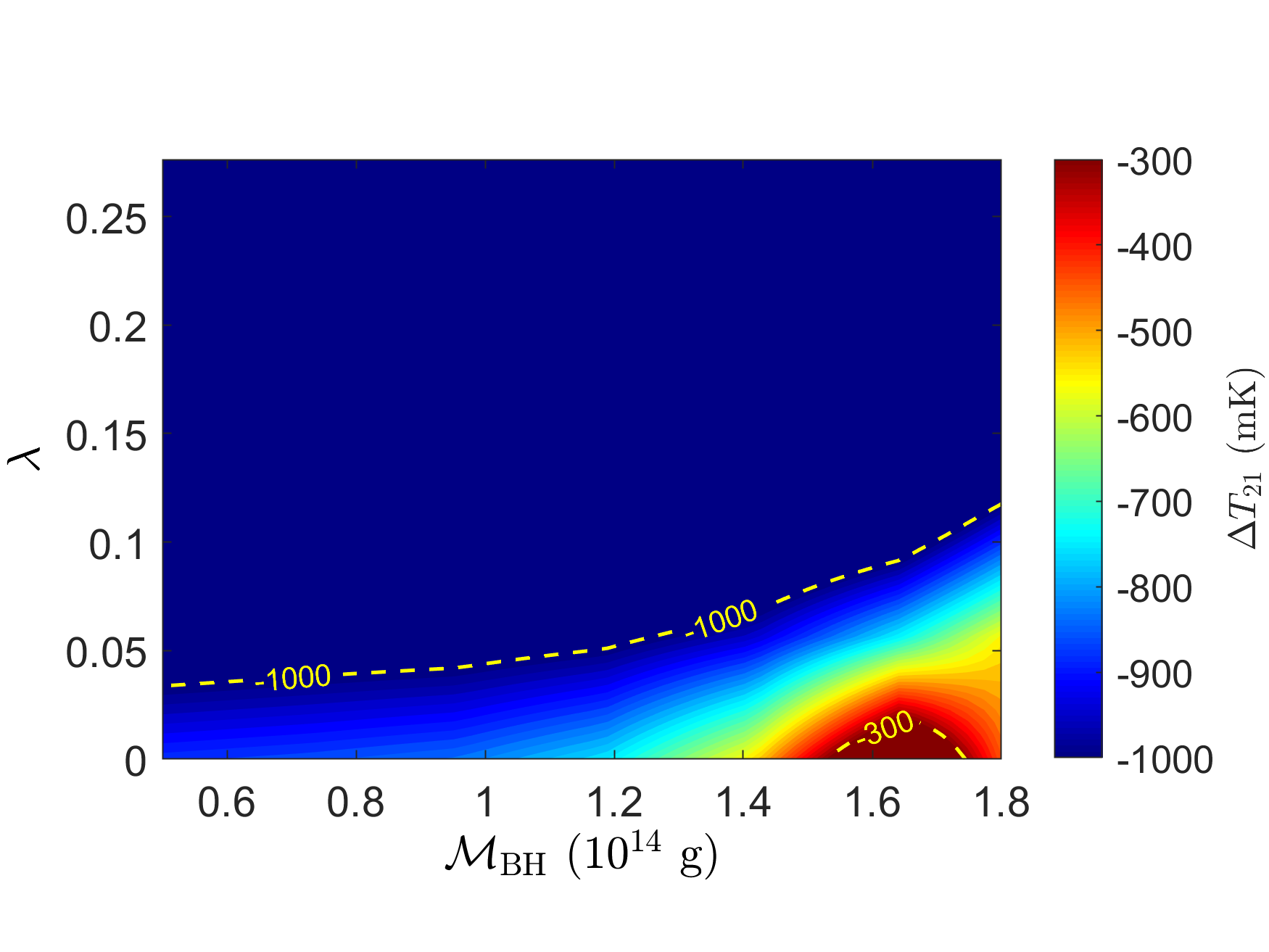}&
		\includegraphics[trim=0 20 80 30, clip, height=0.23\textwidth]{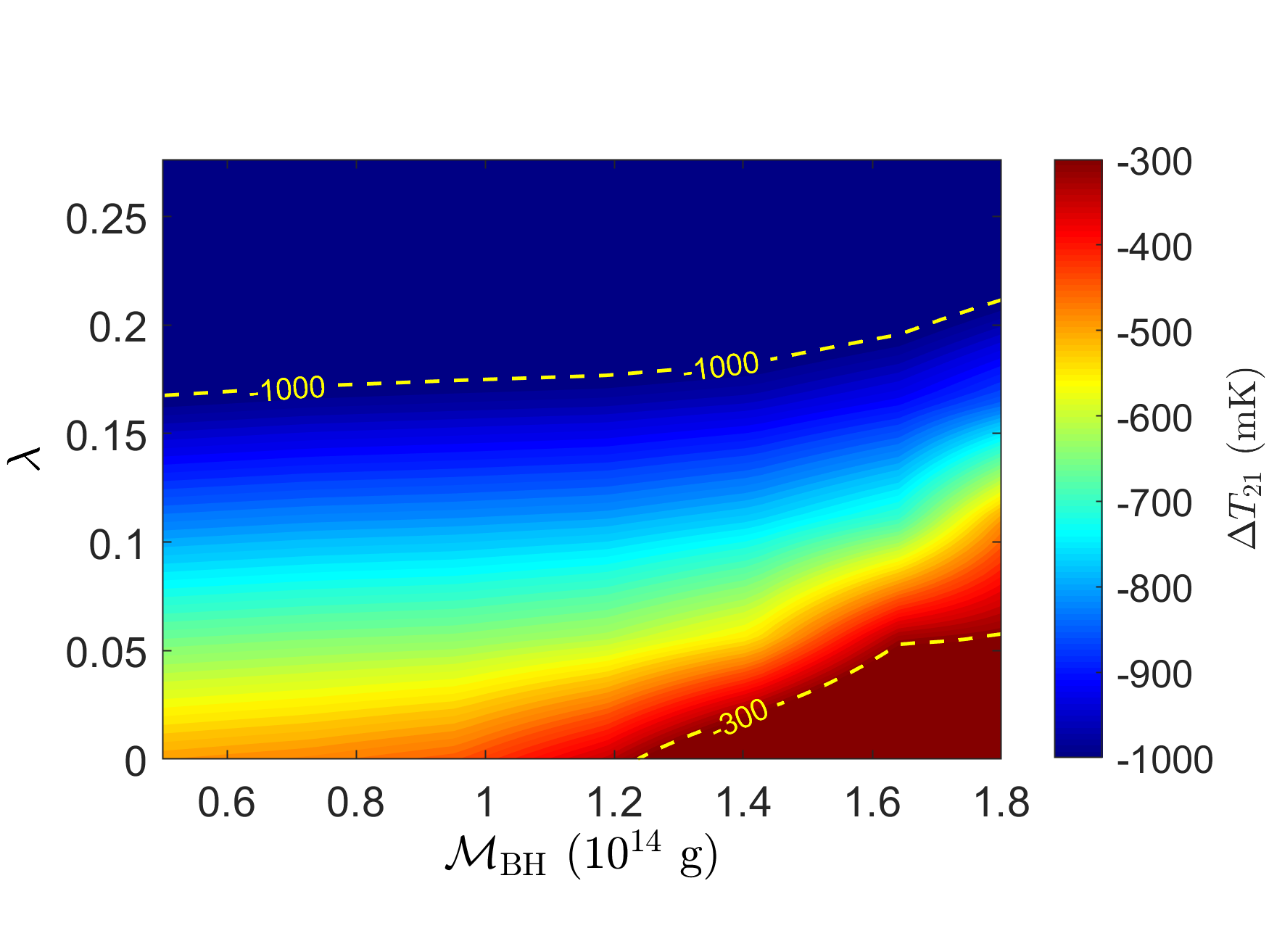}&
		\includegraphics[trim=0 20 0 30, clip, height=0.23\textwidth]{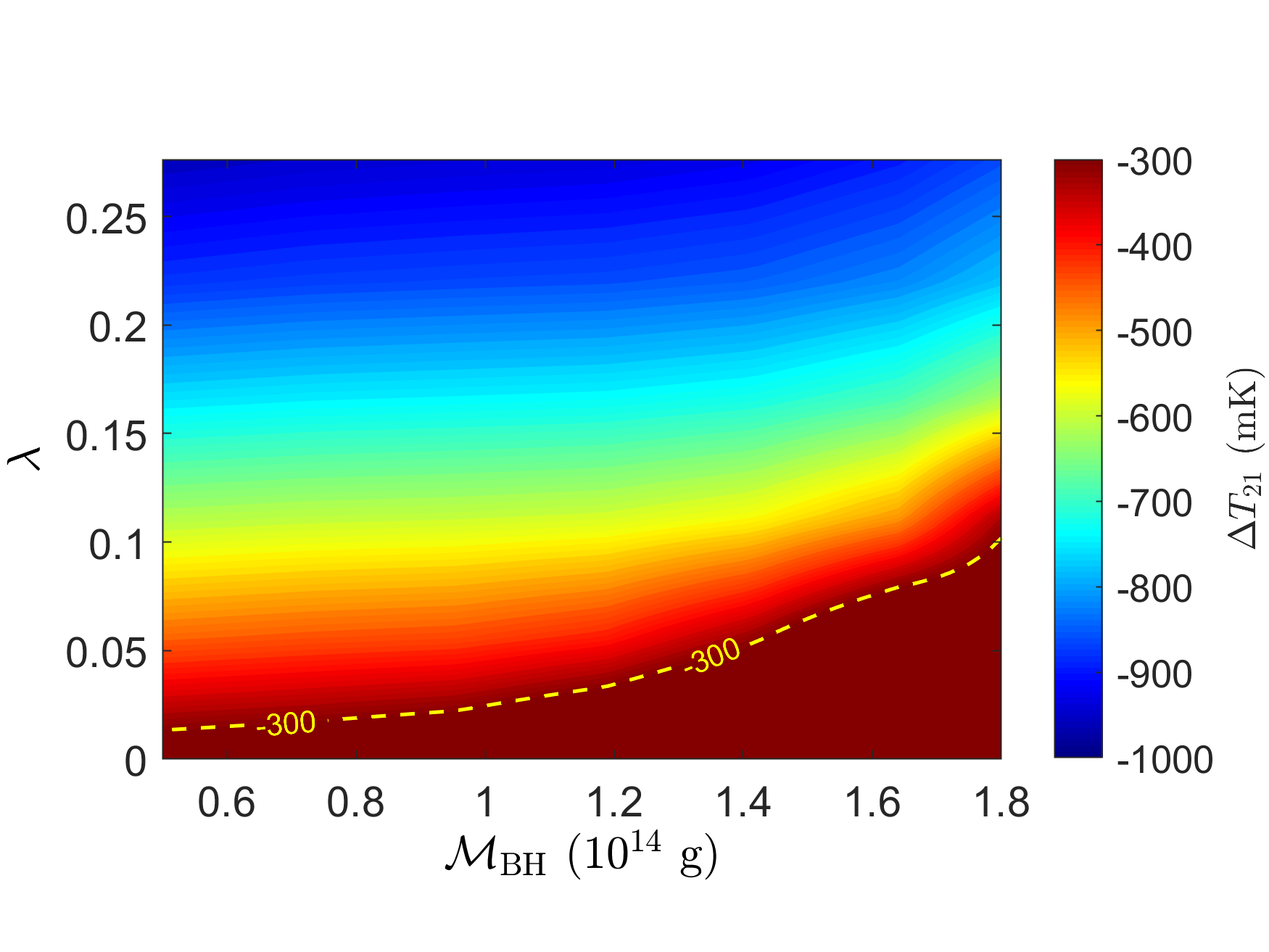}\\
		(g)&(h)&(i)\\
	\end{tabular}
	\caption{\label{fig:mbh_lambda} Variation of $\Delta T_{21}$ with 
		PBH mass and DE-DM coupling parameter $\lambda$ for different DM masses and IDE 
		models (see text for details).}
\end{figure*}

Finally, a detailed study has been carried out to explore similar bounds in the $\mathcal{M}_{\rm BH}$ - $\lambda$ plane for 
individual IDE models and its variation with $m_{\chi}$ as well. We use three different values of dark matter masses namely 
$m_{\chi}=$ 0.5 GeV, 1.0 GeV and 1.5 GeV. In 
each of these cases, the value of $\sigma_{41}$ is set to 1.0. These are shown in Fig.~\ref{fig:mbh_lambda}. The first, second and third row correspond to the 
IDE Model I, II and III respectively while the plots in each of the columns 1, 2 and 3 
are for three fixed dark masses, $m_\chi$ = 0.5 GeV, 1 GeV, 1.5 GeV 
respectively. For example, Fig.~\ref{fig:mbh_lambda}a 
shows the fluctuation of $\Delta T_{21}$ in $\mathcal{M}_{\rm BH}$ - $\lambda$ space for 
$m_{\chi}=0.5$ GeV and the IDE Model I, where the value of 
$\Delta T_{21}$ at each point is described in the colour bar 
shown at the end of 
each row. In all the plots of Fig.~\ref{fig:mbh_lambda}, the yellow dashed 
lines represent the bounds from EDGES result. Fig.~\ref{fig:mbh_lambda}b 
and Fig.~\ref{fig:mbh_lambda}c describe the same for $m_{\chi}=$1.0 GeV and 1.5 GeV 
respectively for IDE Model I. From these figures it can be seen that, 
as $\mathcal{M}_{\rm BH}$ increases, 
$\Delta T_{21}$ increases for a fixed value of $\lambda$ (except 
a little distortion at $\mathcal{M_{\rm BH}}=1.7 \times 10^{14}$g). It can also be noticed  
(from Fig.~\ref{fig:mbh_lambda}a, \ref{fig:mbh_lambda}b 
and \ref{fig:mbh_lambda}c) that as $m_{\chi}$ increases 
(from column 1 to column 3), $\Delta T_{21}$ also increases. 
As mentioned, results for Model II and Model III are furnished 
in plots~\ref{fig:mbh_lambda}d, \ref{fig:mbh_lambda}e, \ref{fig:mbh_lambda}f 
and in plots~\ref{fig:mbh_lambda}g, \ref{fig:mbh_lambda}h, \ref{fig:mbh_lambda}i respectively.
From these figures (all plots of Fig.~\ref{fig:mbh_lambda}), it can be 
noticed that, when 
$m_{\chi}=$ 0.5 GeV, $\lambda$ values for each of the three 
IDE model constraints (see benchmark points described in 
Table~\ref{tab:constraints}) satisfy the EDGES limits for 
$\mathcal{M_{\rm BH}}\lessapprox 1.5\times10^{14}$ g and 
$\mathcal{M_{\rm BH}}\gtrapprox 1.8\times10^{14}$ g. Those 
constraints also agree with the EDGES limit for 
$\mathcal{M_{\rm BH}}\lessapprox 1.2\times10^{14}$ g, when $m_{\chi}=1.0$ is considered. But, for higher values of $m_{\chi}$ ($m_{\chi}=1.5$ GeV), the benchmark values of $\lambda$ 
does not satisfy the EDGES limit of $\Delta T_{21}$ (see 
Fig.~\ref{fig:mchi_sigma}c, Fig.~\ref{fig:mchi_sigma}f and 
Fig.~\ref{fig:mchi_sigma}i).

\section{Summary and Discussions} \label{sec:conc} 
In this work, the effect of PBH evaporation on 21-cm EDGES signal is 
principally addressed while taking into consideration other important 
effects arising out of dark matter scattering on baryons and also 
dark matter-dark energy interactions that can possibly influence 
the observed temperature of 21-cm signal during reionization epoch. 
The PBHs can inject energy into the system through their evaporation 
through Hawking radiation. In addition the dark matter scattering 
off baryons also transfers heat to or from the system depending
on the mass of dark matter. Also, the interactions between dark matter
and dark energy can also influence the 21-cm brightness temperature. To this
end three such interacting dark energy (IDE) models are adopted where
non-minimal coupling of two dark sectors namely dark matter and dark energy
is adopted and the energy transfer due to the IDE and dark matter-baryon 
scattering are properly incorporated
in the relevant evolution equations for baryon temperature, dark matter
temperature etc.

The energy injection rate for the PBH evaporations or in other 
words the contribution of PBHs in the form of Hawking radiation is included
in the present work for the computation of evolution 
of baryon temperature. In fact a set of coupled differential equations
involving evolutions of baryon temperature, dark matter temperature,
the background temperature,
ionization fraction, dark matter baryon relative velocity (leading 
to the drag term) etc. are solved computationally to obtain 
the baryon temperature, spin temperature, the optical depth etc. to finally
compute the 21-cm temperature $T_{21}$. It is to be noted that in the 
present calculations for spin temperature $T_s$, the effect of Lyman-$\alpha$
background is also included where Wouthuysen-Field effect is important.  

The effect of PBH on the 21-cm brightness temperature is demonstrated 
in the present work for different PBH masses ($\sim 10^{14} - 10^{15}$ gm)
and initial PBH mass fraction $\beta_{\rm BH}$ (related to black hole 
number density $n_{\rm BH}$). For the reported 21-cm EDGES signal, the 
allowed regions in the parameter space of
dark matter mass and dark matter - baryon 
scattering cross-sections ($m_\chi-\sigma$) for different 
IDE models (adopted in this work) are obtained for variuos PBH 
masses ${\cal {M}_{\rm BH}}$ and $\beta_{\rm BH}$. It is seen that 
heavier PBH masses appear to favour region of higher $m_\chi$ and lower $\sigma_{41}$. 

It is also to be noted that when dark matter-dark energy interactions are
considered, the dark matter density $\rho_\chi$ and dark energy 
density $\rho_{\rm DE}$ do not evolve as $\sim (1+z)^3$ and 
$(1+z)^{3(1+\omega)}$ suggested by standard cosmology. Therefore the 
evolution of expansion rate of the Universe ($H(z)$) is modified due to 
dark matter-dark energy interactions which in turn affects the optical 
depth and spin temperature of the 21-cm transition. The evolution of Hubble 
parameter is also computed in detail in the present work for all 
the three IDE models adopted. 

The dark matter-dark energy interaction parameter $\lambda$ has been probed 
in this work along with the effects of PBH. The upper and lower limits 
of the IDE parameter $\lambda$ for three IDE models (adopted in this work) 
have been investigated for different 
PBH masses (within the ball park of $10^{14}$ gm) for the range of $T_{21}$
given by the EDGES experiment at reionization era. This appears that the 
model constraints described in Table~\ref{tab:constraints} satisfy the EDGES limit for relatively 
lower masses of dark matter ($\leq 1.0$ GeV). 
It can be observed from Fig.~\ref{fig:tchi} that, although the natures of the evolution of dark matter temperature with $z$ are similar for all the IDE Models I, II and III as also for the case of no DM-DE interaction in the redshift region $1000\leq z\lessapprox 30$, the temperature $T_{\chi}$ for IDE Models I and II always lie above the $T_{\chi}$ for Model I and for no DM-DE interaction case. From Fig.~\ref{fig:tchi} it can be seen that, in case of $m_{\chi} \sim$ GeV, the maximum fluctuation in $T_{\chi}$ is $\sim5$ K for different IDE models. Although a very tiny fluctuation appears in the temperature of dark matter fluid due to DM-DE interaction, it may seed a significant contribution in the global structure formation \cite{structurefrm}. From Fig.~\ref{fig:mbh_lambda}, the limit of IDE coupling parameter $\lambda$ for IDE Model I, II and III can be observed. It can be seen that, although the limits of $\lambda$ barely satisfy EDGES's limit for $m_{\chi}=0.5$ GeV and 1.0 GeV for all three IDE models, they do not agree with the EDGES's limit for higher masses of dark matter particles ($m_{\chi}=1.5$ GeV). Models II and III are very tightly constrained by low-$\ell$ CMB spectrum. The dark matter mass of $m_{\chi}\sim0.5$ GeV, 1 GeV are well within the limit of structure formation. 
Future results related 
to 21-cm physics from early Universe would throw more light to all these issues
and the thermal evolution and dynamics of the early Universe.

\section*{Acknowledgements}
One of the authors (A.H.) wishs to acknowledge the support received from St. 
Xavier’s College, Kolkata and the University 
Grant Commission (UGC) of the Government of India, for providing financial support, 
in the form of UGC-CSIR NET-JRF. One of the authors (MP) thanks the DST-INSPIRE 
fellowship  (DST/INSPIRE/FELLOWSHIP/IF160004) grant by DST, Govt. of India.




\bibliography{PUB21} 



\end{document}